\documentclass[a4paper,11pt]{article}
\usepackage{graphicx} 
\usepackage{tikz} 
\usepackage{placeins}
\usepackage{jinstpub} 
\usepackage{subcaption}
\usepackage{lineno}
\usepackage{amsmath}
\usepackage{multirow}
\usepackage{upgreek}
\usepackage{makecell}
\usepackage{booktabs}
\usepackage{threeparttable}
\usepackage{comment}
\usepackage{xcolor}
\usepackage{float}
\usepackage{pgfplots}
\usepackage{hyperref} 
\usepackage[numbers,sort&compress]{natbib}
\title{Investigation and optimization of the deconvolution method for PMT waveform reconstruction}

\author[a]{J.~Tang,}
\author[a]{T.~Xiao,}
\author[a]{X.~Tang,}
\author[a,1]{and Y. Huang, \note{Corresponding author.}}
\affiliation[a]{School of Physical Science and Technology, Guangxi University, Nanning 530004, China.}

\emailAdd{huangyb@gxu.edu.cn} 

\abstract{Photomultiplier tubes (PMTs) are extensively employed as photosensors in neutrino and dark matter detection. The precise charge and timing information extracted from the PMT waveform plays a crucial role in energy and vertex reconstruction. In this study, we investigate the deconvolution algorithm utilized for PMT waveform reconstruction, while enhancing the timing separation ability for pile-up hits by redesigning filters based on the time-frequency uncertainty principle. This filter design sacrifices signal-to-noise ratio (SNR) to achieve narrower pulse widths. Furthermore, we optimize the selection of signal pulses in the case of low SNR based on Short-Time Fourier Transform (STFT). Monte Carlo data confirms that our optimization yields enhanced reconstruction performance: improving timing separation ability for pile-up hits from $7\sim10$~ns to $3\sim5$~ns, while controlling the residual nonlinearity of charge reconstruction to about 1\% in the range of 0 to 20 photoelectrons.}


\begin{document}
\maketitle
\flushbottom

\section{Introduction}
\label{sec:intro}

Neutrino and dark matter experiments widely use PMT as the detection unit~\cite{SNO:1999crp,FUKUDA2003418,KamLAND:2004mhv,Borexino:2008gab,DayaBay:2016ggj,Cheng2017,DarkSide:2018bpj,AKERIB2020163047}. In order to carry out accurate physical measurements, high-precision energy and vertex reconstruction is indispensable, and they are mainly obtained using accurate PMT charge and hit time information~\cite{Li:2021oos,Yang:2022din,Qian:2021vnh,Zhang:2024okq}. For example, accurate spectral measurements helped the Daya Bay experiment to study the anomaly (“5-MeV bump”) in the reactor neutrino spectrum~\cite{DayaBay:2019yxq}. 

To extract charge and hit time information from PMT waveforms, various commonly used methods have been developed, including simple charge integral method, CR-(RC)$^4$ method, waveform fitting method, and PMT waveform reconstruction methods based on the deconvolution technique~\cite{HUANG201848} or machine learning~\cite{Jiang:2024wph}. The performance and characteristics of several PMT waveform reconstruction methods were summarized in~\cite{HUANG201848}. The simple charge integral and CR-(RC)$^4$ methods are fast and robust. However, in the case of PMT waveforms that exhibit undershoot\footnote{It is called overshoot in the negative signal.} and reflection components due to impedance mismatch in the coupled circuit, the reconstructed charge may be underestimated. This resulting charge nonlinearity poses a challenge for precision energy reconstruction and physical measurements. Additionally, the capability to separate pile-up hits is limited to approximately 20~ns to 40~ns. The deconvolution method was first proposed in the Daya Bay experiment for PMT waveform reconstruction and direct measurement of electronics nonlinearity~\cite{DayaBay:2019fje}, and it has been further applied in experiments such as JUNO~\cite{Zhang_2019,Grassi:2018pxk}. With the deconvolution method, the undershoot was naturally handled, overlapping hits were better separated, and the peak's integral was the number of photoelectrons (PEs). 

In recent years, the advancement of storage and computational resources has prompted some new developments and demands for PMT waveform analysis in neutrino and dark matter experiments. Three of these are briefly summarized as follows: (1) The smearing effect of the PMT charge needs to be improved. For example, the JUNO experiment aims to conduct a precise measurement of energy spectrum derived from reactor neutrinos in order to determine the mass ordering of neutrinos, and the energy resolution is one of the key technical parameters to achieve this goal~\cite{JUNO:2021vlw}. A machine-learning-based PE counting method was proposed to directly predict the number of PEs from PMT waveforms, it can partially mitigate the impact of PMT charge resolution on energy reconstruction and ultimately leading to improved energy resolution~\cite{Jiang:2024wph}. (2) The PMT waveform reconstruction method needs to be applicable for a large dynamic range of several PEs to hundreds of PEs. (3) For the capability to separate pile-up hits, there is still a great potential for improvements. This improvements are expected to be advantageous for particle identification research which utilizes the charge and hit time of each PE~\cite{Chen:2023xhj,Cheng:2023zds}.

In this paper, we propose an enhanced deconvolution method for PMT waveform reconstruction. The enhanced method improves the noise filter and optimizes the selection of signal pulses through STFT. As a result, PMT hits separated by more than $3\sim5$~ns can be discriminated, and the residual nonlinearity of charge reconstruction is controlled to about 1\%. The details of our work will be presented as follows: Section~\ref{sec:sample} introduces the simulation samples of the PMT waveform used in this paper. Then, we review the deconvolution method in section~\ref{sec:deconvolution}. Subsequently, an investigation will be conducted in section~\ref{sec:separation-hit} to explore the influential factors affecting the separation of pile-up hits. In section~\ref{sec:STFT} and section~\ref{sec:Performance}, optimization and performance of the deconvolution algorithm will be shown. Finally, a summary will be provided in section~\ref{sec:summary}

\section{Simulation samples of PMT waveform}
\label{sec:sample}

In this paper, a detailed simulation of the PMT waveform is employed to investigate and validate the performance of the deconvolution algorithm. The model of single photoelectron (SPE) waveform employed in this study was initially developed by Soren JETTER et al. \cite{Jetter_2012}, who parametrically described the waveforms of Hamamatsu R5912 PMTs measured in a laboratory setting. Subsequently, Yongbo Huang et al.~\cite{HUANG201848} utilized measurement data from the FADC system at Daya Bay to enhance and refine the model. As a result, a typical PMT waveform of SPE (denoted as $r(t)$) primarily consists of three components, including the main peak, undershoot, and reflections. In order to facilitate subsequent analysis and comparison, this paper adopts the same PMT waveform model and parameter configuration as presented in~\cite{HUANG201848}. Consequently, equation~\ref{eq:log-norm} can be employed to express the main peak, while equation~\ref{eq:overshoot} can be utilized for expressing the undershoot. The undershoot was parameterized using two exponential functions plus a Gaussian peak. To render the onset of the undershoot, the exponential is multiplied by a Fermi function. The reflections were empirically modeled with the same shape as the main peak, but with different starting time and amplitude (described by parameter $U_{{\rm reflection}}^i$). In the SPE waveform modeling of the Daya Bay experiment, twelve reflection peaks are included, and the time interval between them is 7~ns. As a result, a complete SPE waveform can be described by equation~\ref{eq:sum}, which is also used to model the SPE waveform in our simulation.

\begin{equation}
 \label{eq:log-norm}
 U_{\rm peak}(t) = \begin{cases}
    U^0_{\rm peak} \times \exp\left( -\frac{1}{2} \left( \frac{\ln\left( (t-t_0)/\tau \right)}{\sigma} \right)^2 \right), & \text{if } t > t_0 \\
    0, & \text{if } t \leq t_0
 \end{cases}
\end{equation}

\begin{equation}
 \label{eq:overshoot}
 \begin{split}
 U_{\rm undershoot}\left(t\right) = -U_{\rm peak}\left(t\right)\times & \Bigg[U^{\rm fast}_{\rm undershoot}/\left(e^{\frac{-t+t_{0}+t_{\rm d1}}{\tau_{\rm fermi}}}+1\right)\times e^{\frac{-t+t_{0}-t_{\rm d2}}{\tau_{\rm fast}}} \\
   & + U^{\rm slow}_{\rm undershoot}/\left(e^{\frac{-t+t_{0}+t_{\rm d1}}{\tau_{\rm fermi}}}+1\right)\times e^{\frac{-t+t_{0}-t_{\rm d2}}{\tau_{\rm slow}}} \\
   & + U^{\rm gaus}_{\rm undershoot} \times {\rm exp}\left(-\frac{1}{2}\left(\frac{t-t_{0}-t_{\rm d1}}{\sigma_{\rm undershoot}}\right)^2\right) \Bigg]
 \end{split}
\end{equation}

\begin{equation}
 \label{eq:sum}
 \begin{split}
{U\left(t \right)=U_{{\rm peak}}\left(t \right)+U_{{\rm undershoot}}\left(t \right)+\sum_{i=1}^{12}U_{{\rm reflection}}^i\times U_{{\rm peak}}\left(t-i\times7\right)}
 \end{split}
\end{equation}

\begin{figure}[H]
\centering 
\includegraphics[width=0.55\textwidth]{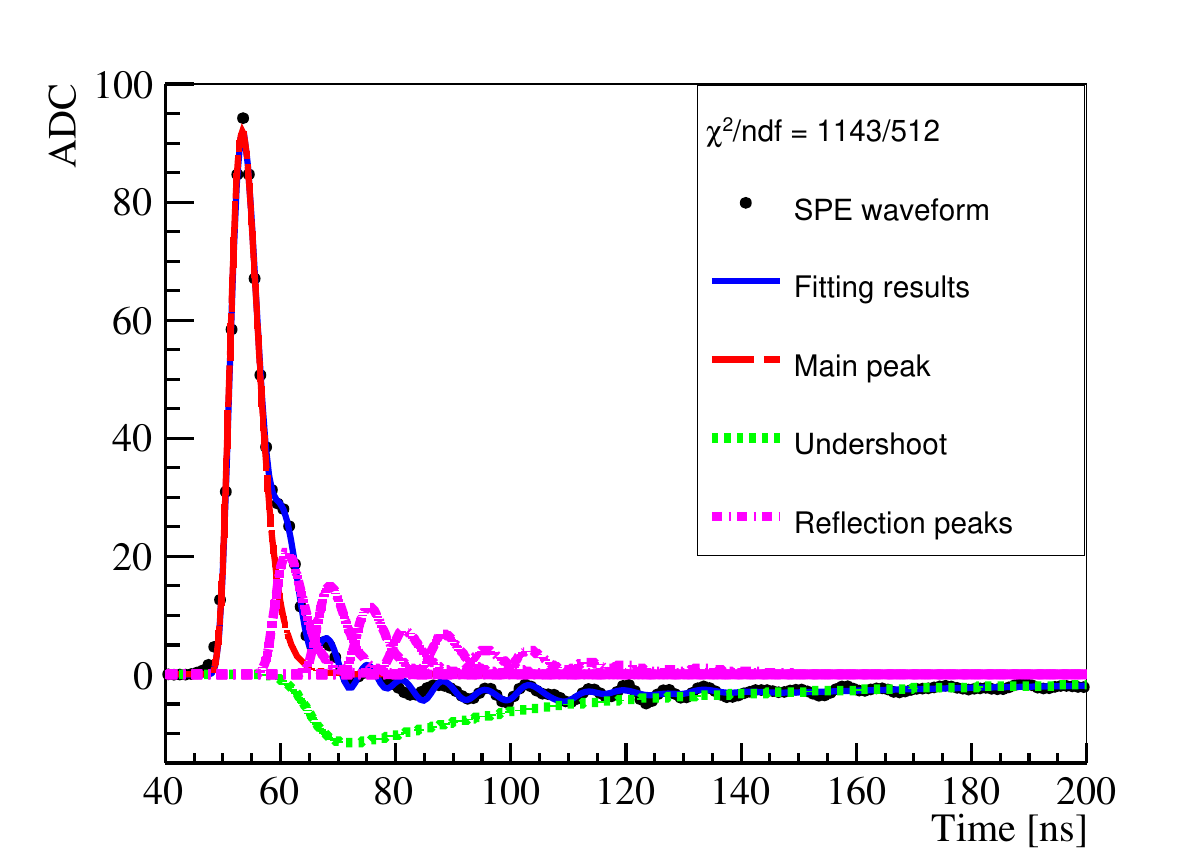}
\caption{\label{fig:SPEWaveform-fitting} SPE waveform of a PMT from Daya Bay data and its fitting result~\cite{HUANG201848}. It consists of three components: the main peak, the undershoot, and twelve reflection peaks. Compared to the main peak, the starting time of the undershoot has a delay of about 20~ns, which can also be found in equation~\ref{eq:overshoot}. The amplitude of the reflections relative to the main peak is approximately 0.24, 0.18, 0.14, 0.1, 0.08, 0.05, 0.02, 0.02, 0.01, 0.008, 0.008 and 0.008, respectively. The relative amplitudes of the three main reflections are also listed in table~\ref{table:parameters}.}
\end{figure} 

\begin{table}[tb]
  \caption{Parameters for the SPE waveform, which are derived from fitting results of Daya Bay FADC data~\cite{HUANG201848}. According to the study of the Daya Bay experiment, the SPE waveform fitting results indicate that some parameters approximately follow a Gaussian distribution. These parameters and their Gaussian resolutions are listed in this table and marked with “Gaussian”. The amplitude was relative to the main peak if no unit included.}
  \label{table:parameters}
  \centering
  \begin{tabular}{cc}
    \hline
    \textbf{Parameters} & \textbf{Values}  \\
    \hline
    Main peak amplitude ($U^0_{\rm peak}$) & 42 mV \\
    Peak width ($\tau$) & 8.4 $\pm$ 0.3 ns (Gaussian) \\
    Peak shape ($\sigma$) & 0.28 $\pm$ 0.02 (Gaussian) \\
    Fast undershoot amp. ($U^{\rm fast}_{\rm undershoot}$) & 0.11 $\pm$ 0.02 (Gaussian)\\
    Slow undershoot amp. ($U^{\rm slow}_{\rm undershoot}$) & 0.03 $\pm$ 0.005 (Gaussian)\\
    $\tau$ for Fermi function ($\tau_{\rm fermi}$) & 10 ns \\
    Fast undershoot ($\tau_{\rm fast}$) & 45 ns\\
    Slow undershoot ($\tau_{\rm slow}$) & 290 ns\\
    Delay time constant of undershoot ($t_{\rm d1}$) & 20 ns\\
    Delay time constant of undershoot ($t_{\rm d2}$) & 5 ns\\
    Gaus undershoot amp. ($U^{\rm gaus}_{\rm undershoot}$) & 0.4 $\times$ $U^{\rm fast}_{\rm undershoot}$\\
    Gaus undershoot shape ($\sigma_{\rm undershoot}$) & 20 ns \\
    1$^{st}$ reflection peak amp. $(U_{{\rm reflection}}^1 )$ & 0.24 $\pm$ 0.02 (Gaussian)\\
    2$^{nd}$ reflection peak amp. $(U_{{\rm reflection}}^2 )$ & 0.18 $\pm$ 0.02 (Gaussian)\\
    3$^{rd}$ reflection peak amp. $(U_{{\rm reflection}}^3 )$ & 0.14 $\pm$ 0.015 (Gaussian)\\
    \hline
  \end{tabular}
\end{table}

An example of an SPE waveform fitting for Daya Bay data is illustrated in figure~\ref{fig:SPEWaveform-fitting}, where the waveform is accurately modeled using equation~\ref{eq:sum}. Detailed descriptions and settings of the SPE waveform parameters are provided in table~\ref{table:parameters}, which were obtained by fitting a large number of SPE waveforms to the FADC system at Daya Bay. We will utilize these parameters in the subsequent waveform simulation of our study. The steps included in the PMT waveform simulation are as follows:

(1) Sample the hit number and the time of occurrence ($t_0$) for each hit. At this stage, each hit corresponds to 1 PE. In our simulation, a total of 200,000 waveforms were generated for subsequent analysis, of which 195,000 waveforms had a hit number uniformly distributed from 0 to 20~PEs, and another 5,000 waveforms had a hit number of 1~PE to ensure statistics in small PE ranges. The time of each hit is sampled according to the fluorescence time profile of liquid scintillator. Two time components are utilized, with respective time constants and weights of 7~ns (80.5\%) and 31~ns (19.5\%).

\begin{figure}[H]
\centering
\begin{subfigure}{0.49\textwidth}
\includegraphics[width=0.99\linewidth]{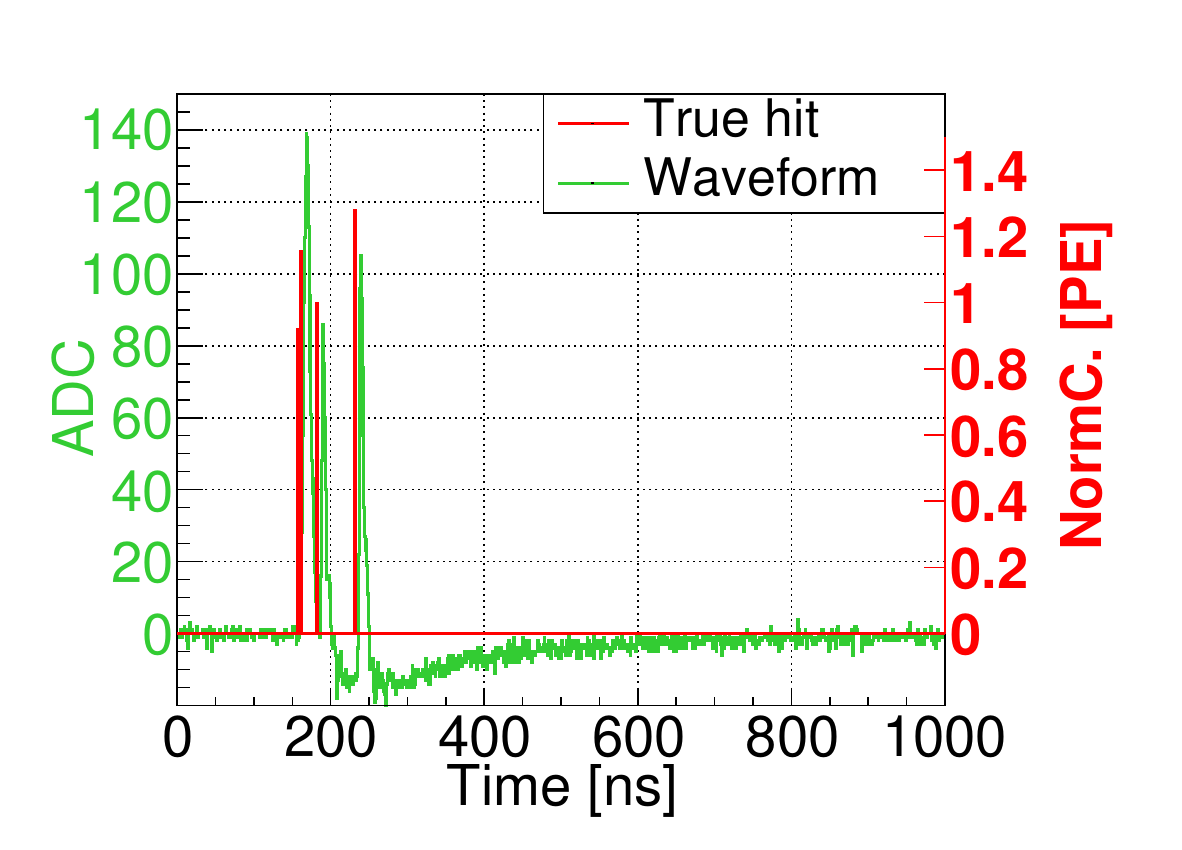}
\caption{\label{fig:Waveform_Charge:a}Full waveform.}
\end{subfigure}
\begin{subfigure}{0.49\textwidth}
\includegraphics[width=0.99\linewidth]{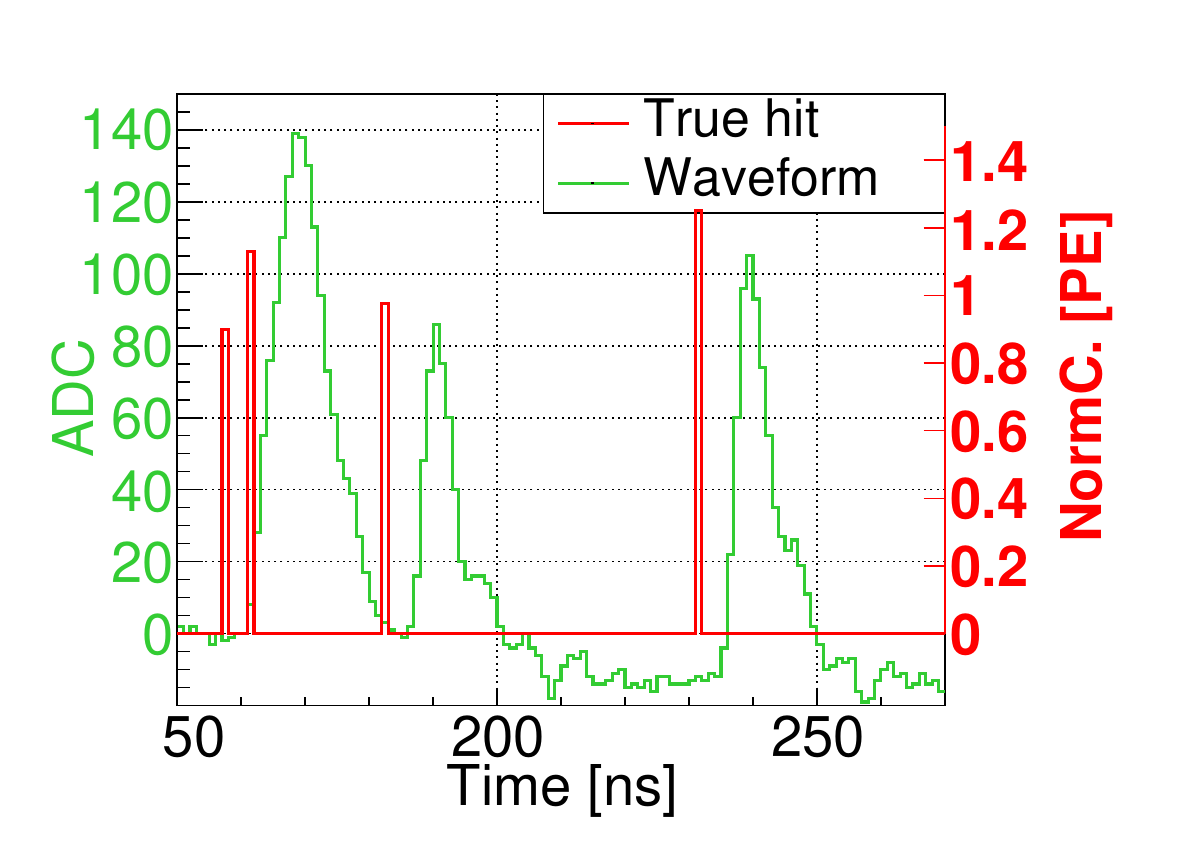}
\caption{\label{fig:Waveform_Charge:b} Partial waveform of figure~\ref{fig:Waveform_Charge:a}.}
\end{subfigure}
\caption{\label{fig:Waveform_Charge} An example of a PMT waveform (shown in green) that corresponds to four hits (marked as “True hit” and shown in red), exhibiting a distinct undershoot component. The full waveform is shown in (a), while the narrower range (50-270~ns) is shown in (b). In the figure, the amplitude of each hit is denoted as the normalized charge (“NormC. [$\rm PE$]”), and the area of each hit indicates its charge. Thus, the charges of these four hits are 0.91~PEs, 1.13~PEs, 0.97~PEs and 1.25~PEs, respectively.}
\end{figure}

(2) The amplitude of each hit is diffused with a Gaussian distribution, and the Gaussian sigma is set to 30\%, which corresponds to the charge resolution of a SPE in PMT. The example depicted in figure~\ref{fig:Waveform_Charge} illustrates four sampled hits occurring at 157~ns, 161~ns, 182~ns, and 231~ns respectively. After taking into account the charge dispersion of individual photoelectrons, their charges are 0.91~PEs, 1.13~PEs, 0.97~PEs and 1.25~PEs, respectively. After SPE charge smearing, the PMT hit combinations that carry charge and time information are referred to as “true hits” ($u(t)$) in this paper. These true hits correspond to the red color in figure~\ref{fig:Waveform_Charge}, and each of them has a width of 1~ns, corresponding to a sampling rate of 1~GHz (described in Step (5)). The integral of the amplitude of each hit within its width (i.e., the area of the hit) represents the charge of the hit. In the figure, the amplitude of each hit is denoted as the normalized charge (“NormC. [$\rm PE$]”).

(3) Convolve true hit $u(t)$ with analog PMT SPE waveform $r(t)$, which was introduced at the beginning of section~\ref{sec:sample} and generated using equation~\ref{eq:sum}. $r(t)$ is also known as the convolution kernel.

(4) Add Gaussian white noise $n(t)$ with a sigma of 0.7~mV into the analog PMT waveform to simulate analog electronics noise. Therefore, a measured waveform $m(t)$ can be describe as equation~\ref{eq:waveform}.

\begin{equation}
\label{eq:waveform}
m\left(t \right) = u\left(t \right) \ast r\left(t \right) + n\left(t \right)
\end{equation}

(5) The analog waveform was digitized using the same FADC configurations as Daya Bay. The length of each waveform is 1000~ns, with a sampling rate of 1~GHz. Consequently, the interval between two sampling points in the waveform is 1~ns. The SPE amplitude is about 40-50~mV after 10$\times$ amplification in the Daya Bay FADC system, depending on the specific operating high voltage of each PMT. This amplitude corresponds to about 80-110 ADC counts, with approximately 2.2 ADC counts per mV. This paper refers to the SPE waveform modeling and simulation in~\cite{HUANG201848}, as a result, the amplitude of an SPE waveform ($\sim$42~mV) corresponds to approximately 94~ADC. An example of the simulated waveform is shown in figure~\ref{fig:Waveform_Charge}. In precise measurements, the waveform is expected to be reconstructed to 4.26 PEs and it actually corresponds to a true photoelectron number of 4 PEs before the dispersion effect of the PMT measurement.

\section{The deconvolution method for PMT waveform reconstruction}
\label{sec:deconvolution}

\begin{figure}[H]
\centering 
\includegraphics[width=0.49\textwidth]{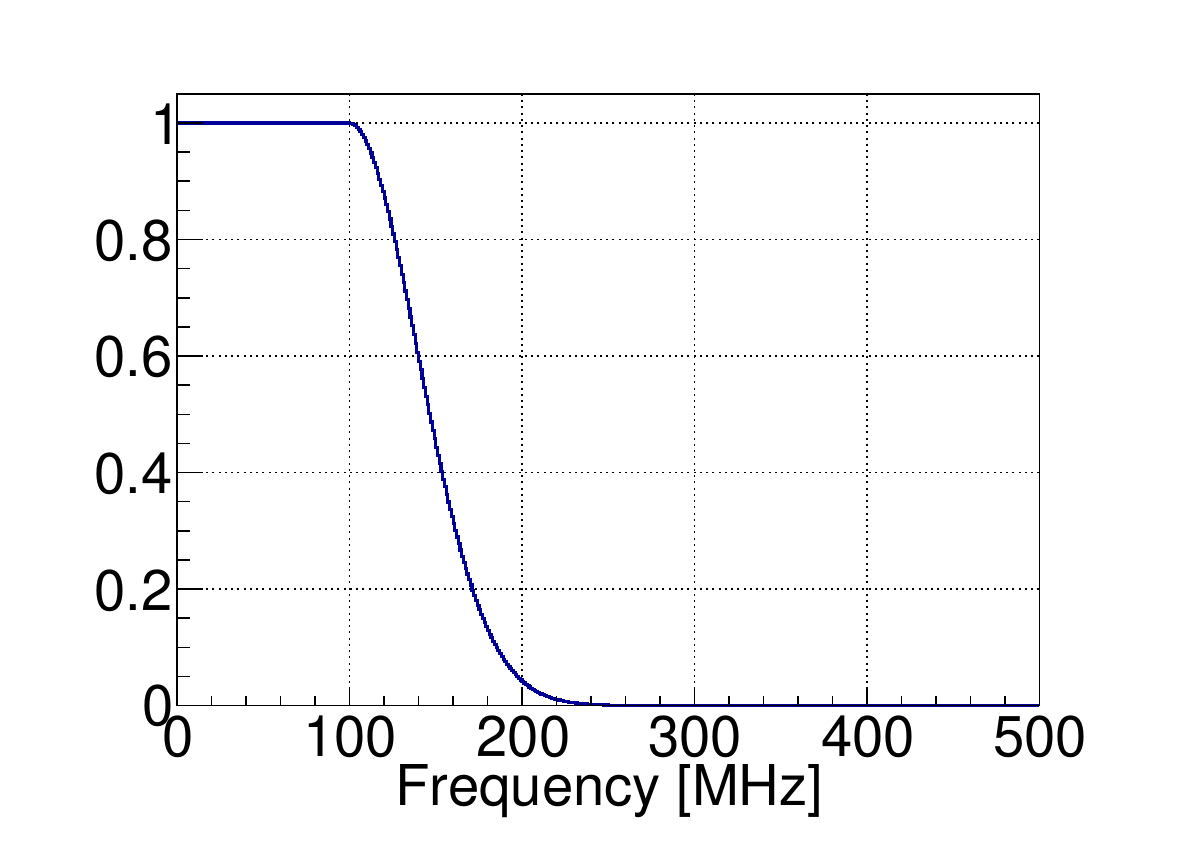}
\caption{\label{fig:FFT_GausFilter_old} A custom-defined low-pass filter. The filter response equals 1 for frequencies less than 100 MHz, and is 0 for frequencies greater than 300 MHz. Between 100 MHz and 300 MHz, the filter is described by the Gaussian formula, whose mean and standard deviation (sigma) values are 100 and 40, respectively.}
\end{figure}

\begin{figure}[H]
\centering
\begin{subfigure}{0.49\textwidth}
\includegraphics[width=0.99\linewidth]{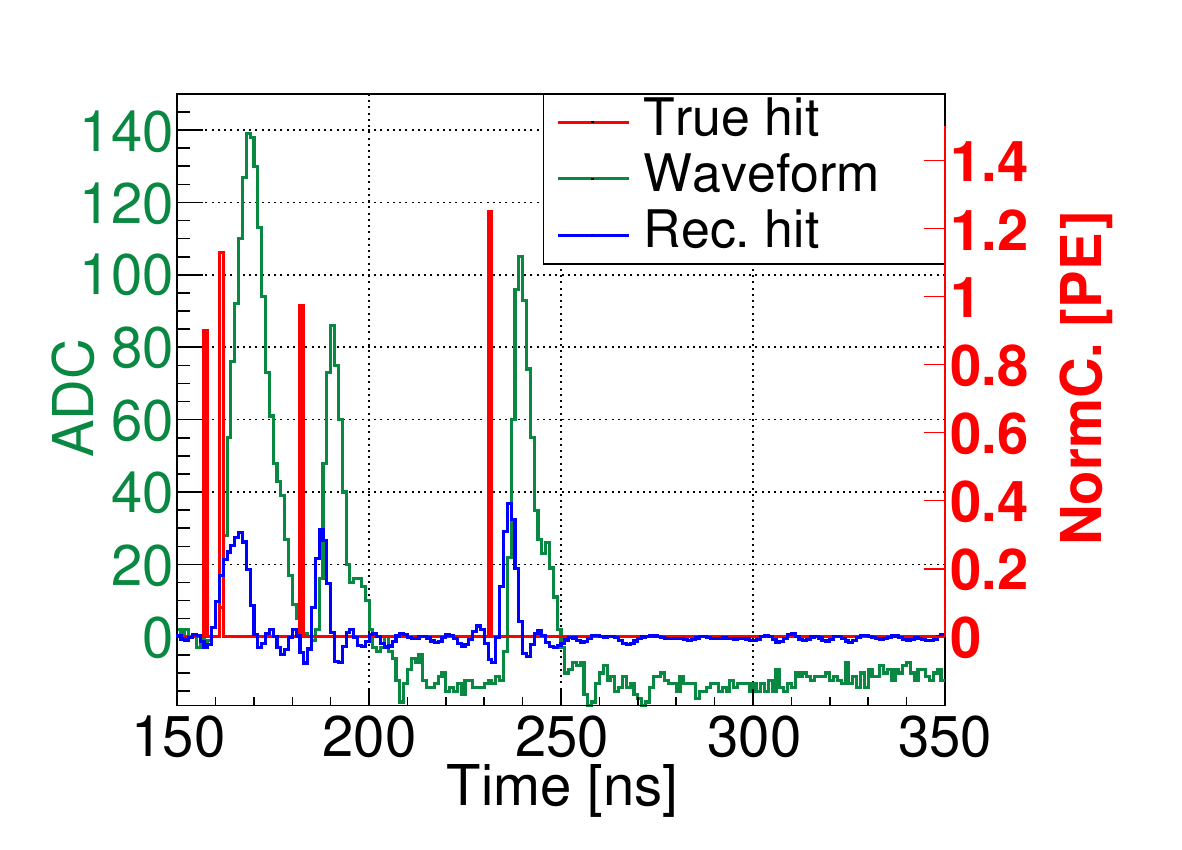}
\caption{\label{fig:FFT_Guas_back:a}Reconstructed result.}
\end{subfigure}
\begin{subfigure}{0.49\textwidth}
\includegraphics[width=0.99\linewidth]{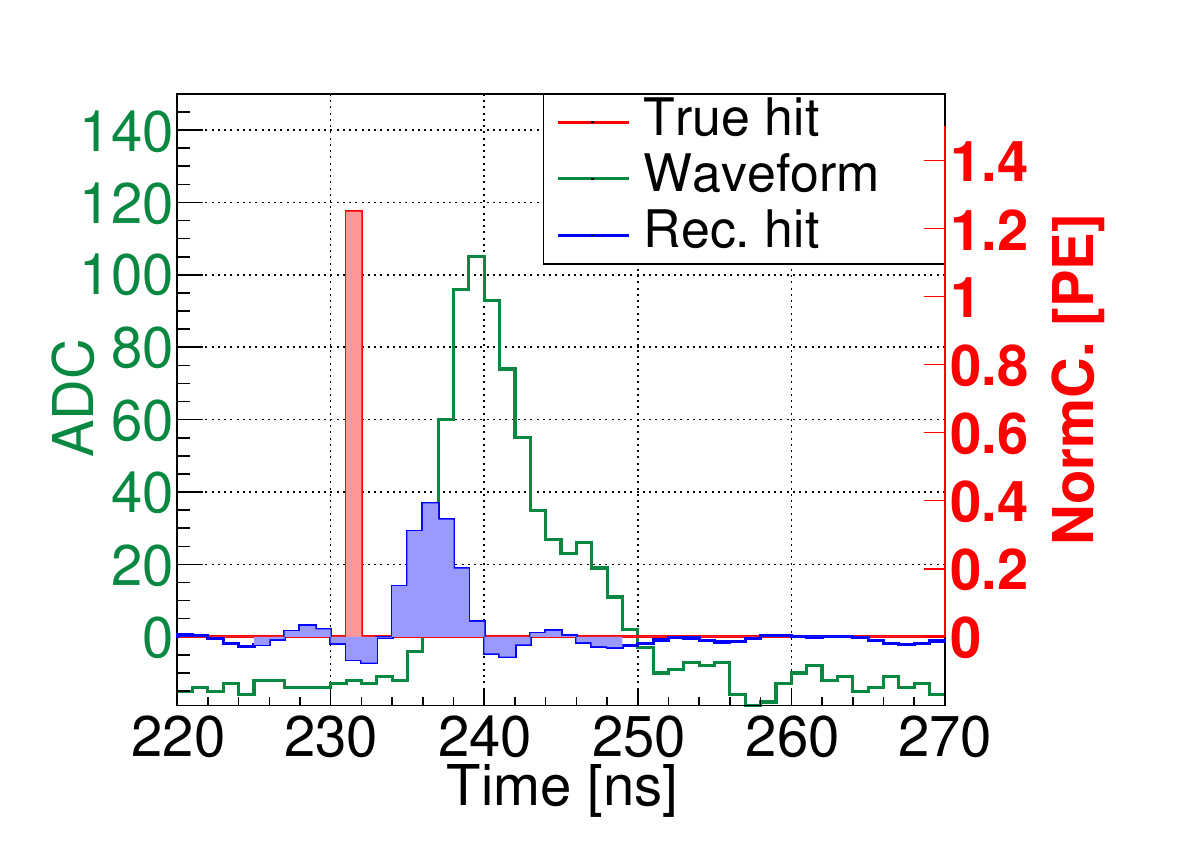}
\caption{\label{fig:FFT_Guas_back:b}Local details of figure~\ref{fig:FFT_Guas_back:a} in the 220-270~ns range.}
\end{subfigure}
\caption{\label{fig:FFT_Guas_back} Reconstruction results corresponding to the waveform in figure~\ref{fig:Waveform_Charge}. The blue peaks correspond to the reconstructed hits, which is narrower than the original waveform (shown in green) but wider than the true hits (shown in red), and the undershoot has been recovered. The details of figure~\ref{fig:FFT_Guas_back:a} in the range 220-270~ns can be found in figure~\ref{fig:FFT_Guas_back:b}. The blue shade represents the integral region for charge calculation, encompassing the peak region of the reconstructed hit, as well as the 9~ns interval preceding the left edge of the peak and the 9~ns interval following the right edge of the peak.}
\end{figure}

The primary consideration of the deconvolution algorithm for PMT waveform reconstruction is to directly extract the true hits ($u(t)$) and calculate the corresponding charges, thereby avoiding underestimation issues when reconstructing waveforms with complex components such as undershoot. First, the time domain waveform of PMT is converted to the frequency domain with Discrete Fourier Transform (DFT), and a low-pass filter  (an example is shown in figure~\ref{fig:FFT_GausFilter_old}) is adopted to remove the high-frequency electronic noise, then divided by the frequency response of the calibrated SPE waveform ($r_{\rm calib}(t)$) to make the reconstruction insensitive to waveform characteristics. Next, the reconstructed hits ($u_{\rm rec}(t)$) can be obtained after converted to the time domain with Inverse DFT. The DFT and Inverse DFT operations in this paper are based on the Fast Fourier Transform package in ROOT~\cite{BRUN199781}. The above calculation steps can be described by equation~\ref{eq:deconv_steps}. The reconstructed hits ($u_{\rm rec}(t)$) corresponding to the waveform in figure~\ref{fig:Waveform_Charge} are depicted in figure~\ref{fig:FFT_Guas_back:a}, demonstrating effective management of undershoot and improved separation of pile-up hits.

\begin{equation}
\label{eq:deconv_steps}
\begin{gathered}
u_{\rm rec}\left(t \right) = \mathcal{F}^{-1}\left[\frac{\mathcal{F}[m\left(t \right)] \times filter\left(f \right)}{\mathcal{F}\left[r_{\rm calib}\left(t \right) \right]}\right]
\end{gathered}
\end{equation}

In the reconstruction, we applied a custom-defined low-pass filter as shown in figure~\ref{fig:FFT_GausFilter_old}, and it was designed based on the experience gained from FADC waveform reconstruction in the Daya Bay experiment. Low-pass filters are commonly employed to effectively attenuate high-frequency noise while preserving the low-frequency component of the signal, thereby significantly enhancing the SNR. The Gibbs effect~\cite{GibbsEffect}, on the other hand, causes local ringing near the peaks due to high-frequency cutting of the noise filter. Finally, reconstructed charge can be achieved by selecting an appropriate integral region, as demonstrated in~\cite{HUANG201848}. In this figure, the area of each reconstructed hit (blue) is nearly the same as the area of its corresponding true hit (red), indicating that the charge reconstruction was successful. Typically, PMT waveform reconstruction using the deconvolution method described above can separate hits with a time interval greater than 7-10~ns, while controlling the charge nonlinearity within 1\%. 

\section{Investigation and optimization of the deconvolution method}
\label{sec:Investigation-optimization}

In this paper, the pulse width of a single hit is defined as the time interval between the rising and falling edges of the hit that is consistently higher than the baseline. If two hits occur simultaneously, they will completely overlap in time; however, this is not the subject of this study. The relevant analysis will be further discussed in the near future. This paper mainly discusses the optimal capability of the deconvolution algorithm to distinguish between two hits when there is a non-zero time interval, i.e., it is primarily related to the pulse width of the reconstructed hit. Therefore, this paper mainly uses the pulse width of a single hit to characterize the algorithm's ability to separate pile-up hits. In the reconstruction results of section~\ref{sec:deconvolution}, it can be found that the reconstructed hit exhibits a broader pulse width (width$\sim$7~ns) compared to the true hit (width = 1~ns), thereby imposing limitations on further separation of pile-up hits, and the local ringing will also interfere with the charge calculation. In this section, we will investigate the factors that contribute to the additional broadening of the reconstructed hit and explore the potential of using deconvolution algorithms to enhance the separation of pile-up hits.

\subsection{Factors influencing the separation of pile-up hits}
\label{sec:separation-hit}
This section investigates the two primary factors that contribute to pulse width in PMT waveform reconstruction using the deconvolution algorithm, including convolution kernel mismatch and the effects of the low-pass filter.

\subsubsection{Convolution kernel mismatch}
\label{sec:Fluctuations}

\begin{figure}[H]
\centering 
\includegraphics[width=0.55\textwidth]{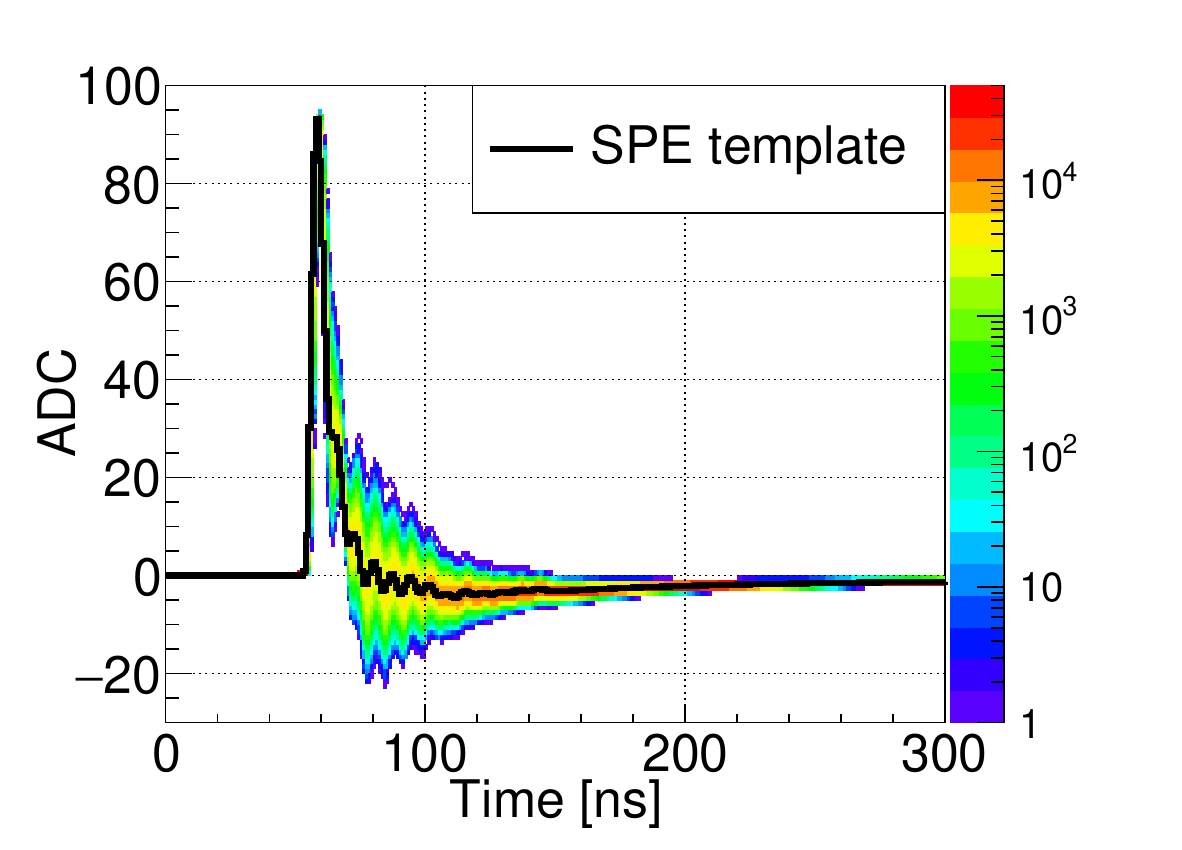}
\caption{\label{fig:SPE-fluctuation} The SPE waveform template and 10000 SPE waveforms after shape fluctuation. These fluctuations are based on the waveform measurement data of the Daya Bay FADC system. The effect of PMT on waveform fluctuation is shown, where the wobble of reflection and undershoot is relatively large. The fluctuation of waveform shape is an important factor affecting deconvolution performance.}
\end{figure}

The deconvolution algorithm, as described in section~\ref{sec:deconvolution}, requires a convolution kernel (the calibrated SPE waveform $r_{\rm calib}(t)$). The convolution kernel can be obtaind by using the data for SPE calibration and then selecting SPE waveforms for averaging. However, as shown in table~\ref{table:parameters}, there are slight fluctuations in the SPE waveform shape of a PMT, indicating that there will always be some mismatch when employing the frequency response of the calibrated SPE waveform as the convolution kernel for reconstructing other waveforms. Ten thousand SPE simulated waveforms with a charge of 1 PE and shapes fluctuating based on the parameters in table~\ref{table:parameters} are displayed in figure~\ref{fig:SPE-fluctuation}. The black waveform corresponds to the waveform obtained using the central values of these parameters and is also referred to in this article as the SPE waveform template. 

In order to facilitate the separate discussion of convolutional kernel and filter effects, we replaced $m(t)$ in equation~\ref{eq:deconv_steps} by equation~\ref{eq:waveform} and obtained:

\begin{equation}
\label{eq:deconv_details}
\begin{gathered}
u_{\rm rec}\left(t \right) = \mathcal{F}^{-1}\left[\mathcal{F}[u\left(t \right)] \times filter\left(f \right) \times \frac{\mathcal{F}\left[r\left(t \right) \right]}{\mathcal{F}\left[r_{\rm calib}\left(t \right) \right]} + filter\left(f \right) \times \frac{\mathcal{F}\left[n(t)\right]}{\mathcal{F}\left[r_{\rm calib}\left(t \right) \right]} \right]
\end{gathered}
\end{equation}

\begin{figure}[H]
\centering
\begin{subfigure}{0.49\textwidth}
\includegraphics[width=0.9\linewidth]{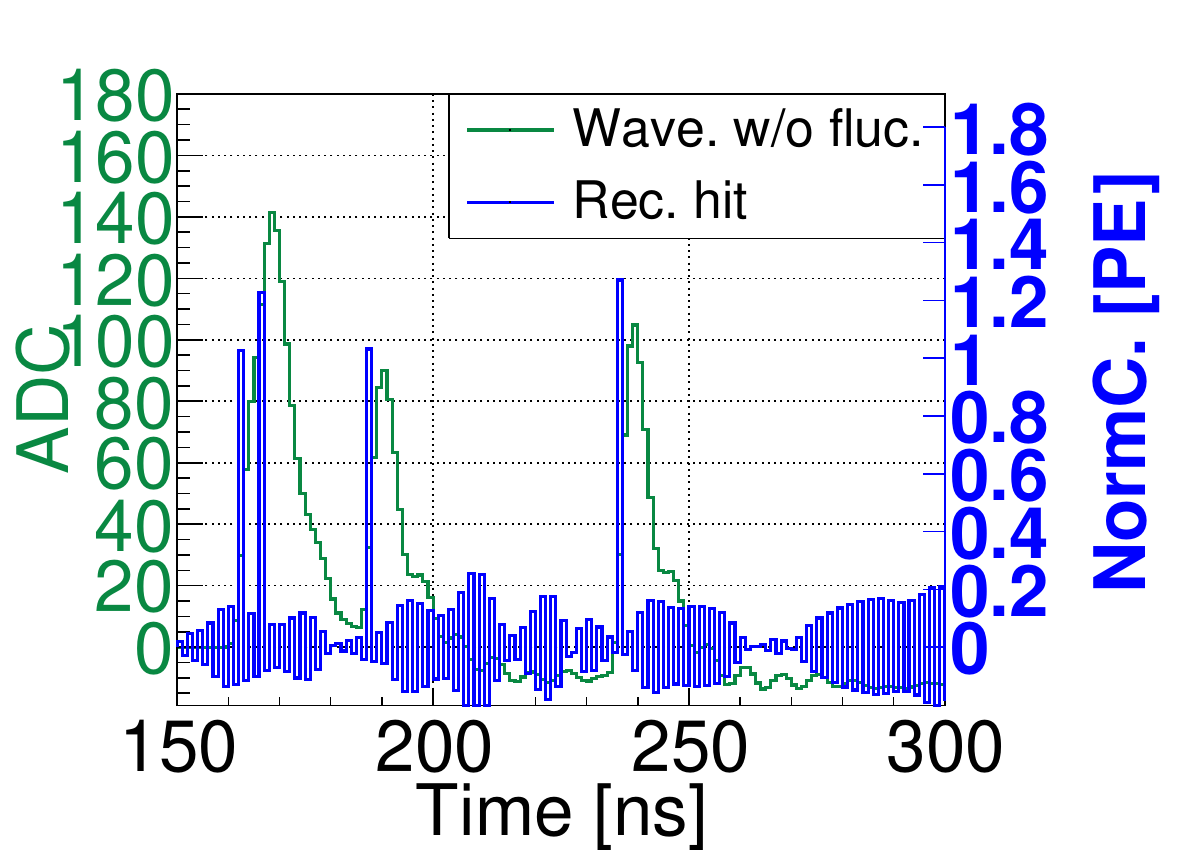}
\caption{\label{fig:FFT_match}SPE waveform without fluctuation.}
\end{subfigure}
\begin{subfigure}{0.49\textwidth}
\includegraphics[width=0.9\linewidth]{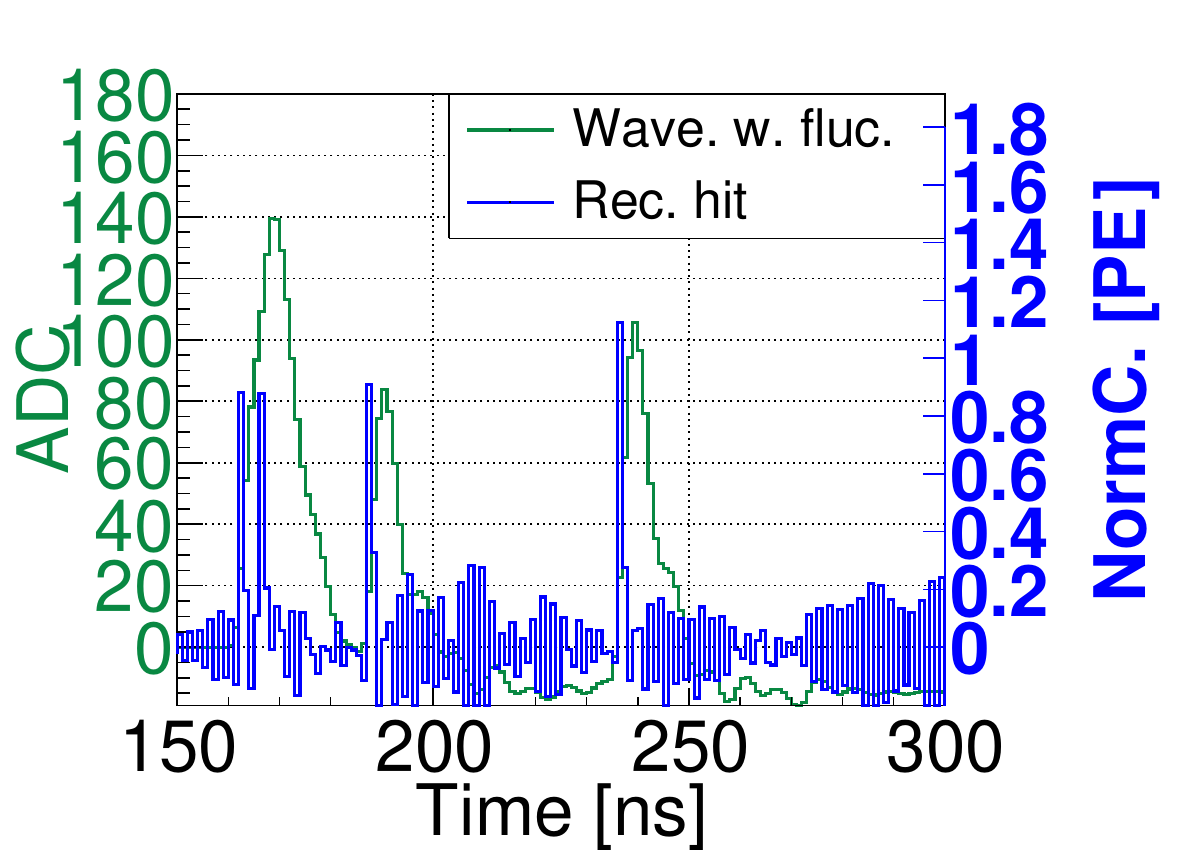}
\caption{\label{fig:fluctuation}SPE waveform with fluctuation.}
\end{subfigure}
\caption{\label{fig:FFT_match-AND-fluctuation} Waveform reconstruction results without (figure~\ref{fig:FFT_match}) and with (figure~\ref{fig:fluctuation}) shape fluctuations. For comparison purposes, we still use the case of 4 true hits occurring at 157~ns, 161~ns, 182~ns, 231~ns as shown in figure~\ref{fig:Waveform_Charge}. However, no electronic noise is incorporated into the simulation, and no noise filter is applied during the reconstruction. Shape fluctuation is disabled for figure~\ref{fig:FFT_match}, whereas it is enabled for figure~\ref{fig:fluctuation}. The results show that when the convolution kernel is matched, the reconstructed hits (shown in blue) with pulse width of 1~ns are obtained. Instead, the pulse width increases slightly to 2~ns. The high-frequency oscillation is due to the fact that the baseline in the original waveform (shown in green) is not fully restored, specifically referring to the text description. Note that figure~\ref{fig:FFT_match} does not directly illustrate the baseline failing to recover to zero by the end of the time window (i.e., the baseline value should be 0~ADC at t = 1000~ns), as the figure only displays the waveform up to 300~ns. In fact, the baseline value at the end of the time window is -0.82~ADC.}
\end{figure}

To investigate the influence of convolution kernel on the pulse width of reconstructed hits independently, we conducted waveform simulations and reconstructions without incorporating electronic noise ($n(t)=0$) and filtering ($filter(f)=1$). Consequently, in cases where the convolution kernel is perfectly matched ($r(t)=r_{\rm calib}(t)$), each hit is expected to be accurately reconstructed  ($u_{\rm rec}(t)=u(t)$), resulting in a pulse width of 1~ns for each hit reconstructed by deconvolution. However, as shown in figure~\ref{fig:FFT_match}, where we turn off the fluctuation of waveform shape, high-frequency oscillations were present on the baseline in addition to hits with a 1~ns pulse width in the reconstructed results. These high-frequency oscillations are caused by spectral leakage. The FFT assumes periodicity in the signal, and thus requires the signal length to align with the assumed period. However, PMT waveforms are non-stationary~\cite{Spagnolini2017} signals. When applying the DFT~\cite{Sundararajan2024} to a PMT waveform, the algorithm interprets the waveform's length as one period of the cyclic signal. If the waveform does not return to its baseline within the sampling time window, the DFT assumes the signal is connected head-to-tail, leading to the introduction of additional frequency components, which is what we refer to as spectral leakage. The effect is more serious and requires special consideration and treatment for large PE waveforms with significant undershoot. In this study, although spectrum leakage introduces noise into the deconvolution results, it is only a minor factor that does not affect the reconstruction results. Specifically, it can be mitigated by applying an appropriate filter, and information from the STFT can be utilized to accurately select the signal pulse in this effect and perform subsequent reconstruction, which will be discussed in later sections.

When the convolution kernel is mismatched ($r(t) \neq r_{\rm calib}(t)$), the ill-posedness of deconvolution leads to an increased pulse width in the reconstructed hit. When the fluctuation of the SPE waveform shape is set according to table~\ref{table:parameters}, as demonstrated in figure~\ref{fig:fluctuation}, the pulse width of the reconstructed hit increases slightly from 1~ns to 2~ns.

\subsubsection{Low-pass filter}
\label{sec:low-pass filter}

The convolution kernel mismatch, as discussed in section~\ref{sec:Fluctuations}, introduces high-frequency noise into the deconvolution result, while the electronic noise also predominantly exhibits high frequencies. Therefore, it is essential to incorporate a low-pass filter during the reconstruction process. This section mainly discusses the impact of low-pass filters on the performance of the deconvolution reconstruction.

The Time-Frequency Uncertainty Principle states that it is impossible to precisely determine both the time and frequency localization of a signal simultaneously. Specifically, if a signal is highly localized in time, its frequency representation will necessarily be spread out, indicating poor frequency localization. Conversely, if a signal is highly localized in frequency, it must be spread out in time, indicating poor time localization~\cite{Heisenberg1927, Griffiths_Schroeter_2018, Boughn2017, Ream1977}. 

According to the time-frequency uncertainty principle, when a signal has a narrow pulse width in the time domain, it corresponds to a wide spectrum across multiple frequencies. However, when a low-pass filter is adopted, it limits the high-frequency components of the signal due to its limited bandwidth, resulting in an increase in the pulse width in the time domain. This means that using a low-pass filter in the deconvolution algorithm will cause an increase in the pulse width of the reconstructed hit. This is consistent with the time-frequency uncertainty principle.

\begin{figure}[H]
\centering
\begin{subfigure}{0.49\textwidth}
\includegraphics[width=0.95\linewidth]{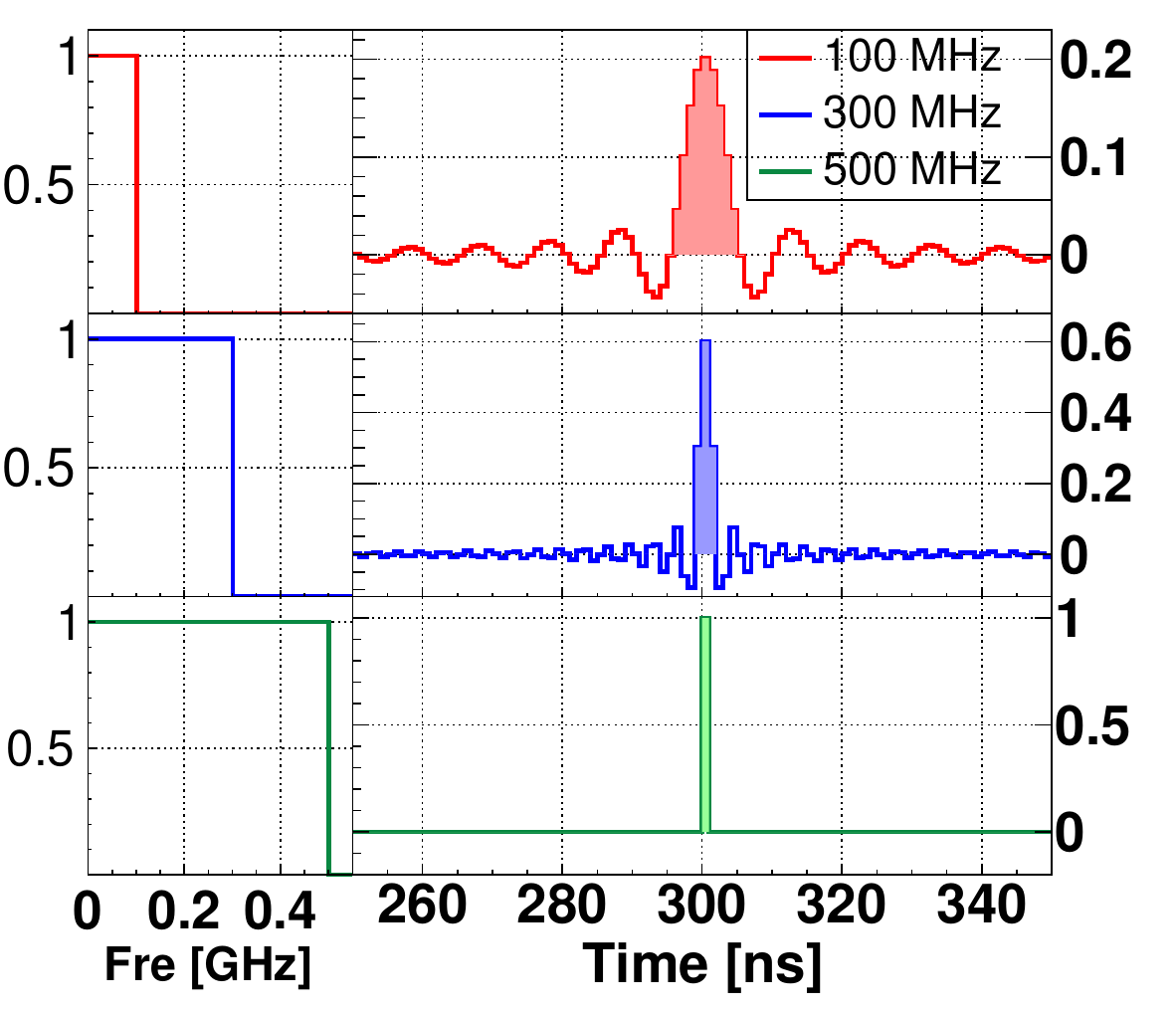}
\caption{\label{fig:ideala:a}The ideal low-pass filter.}
\end{subfigure}
\begin{subfigure}{0.49\textwidth}
\includegraphics[width=0.95\linewidth]{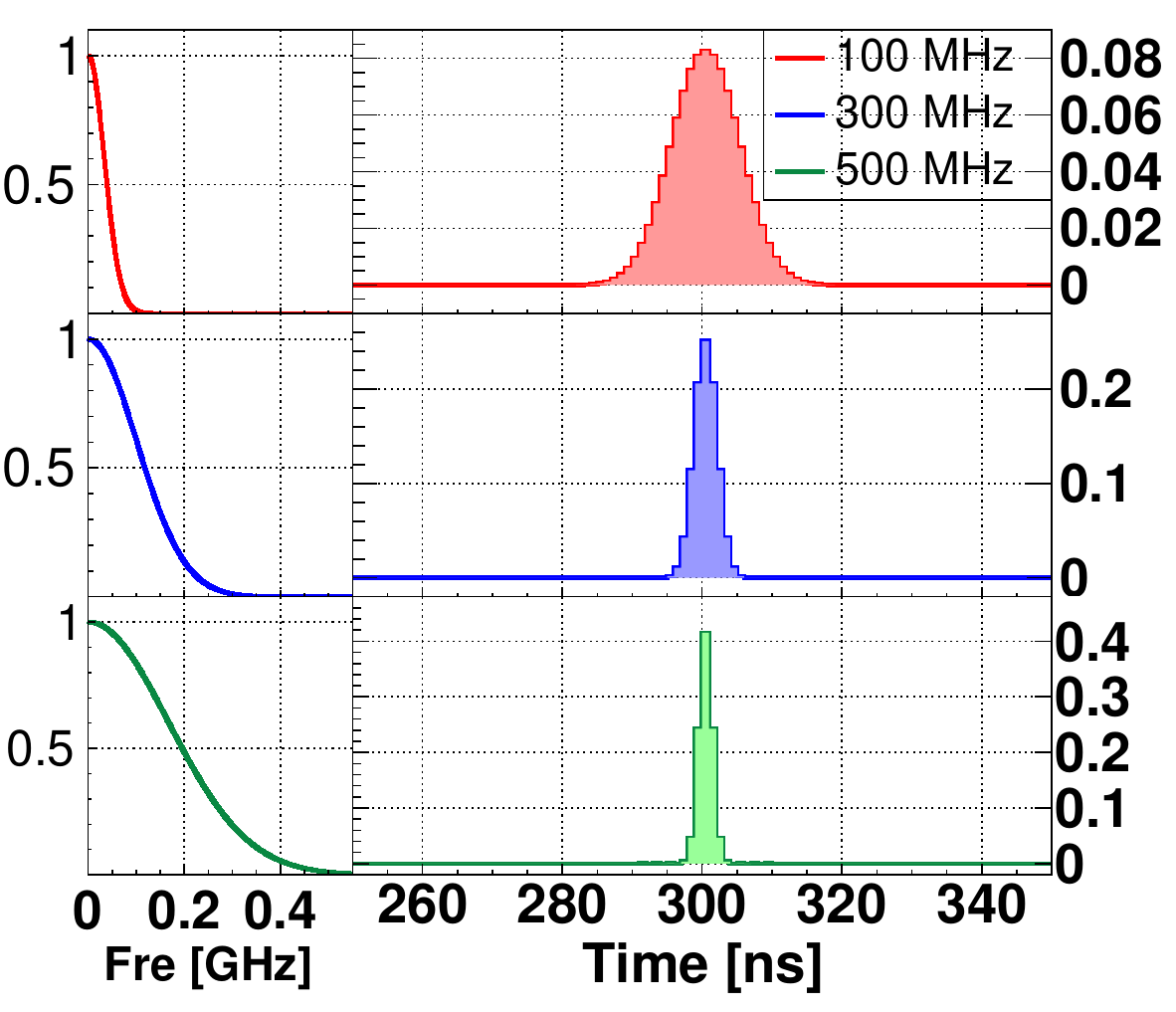}
\caption{\label{fig:gausb:b}The pure Gaussian filter.}
\end{subfigure}
\begin{subfigure}{0.49\textwidth}
\includegraphics[width=0.95\linewidth]{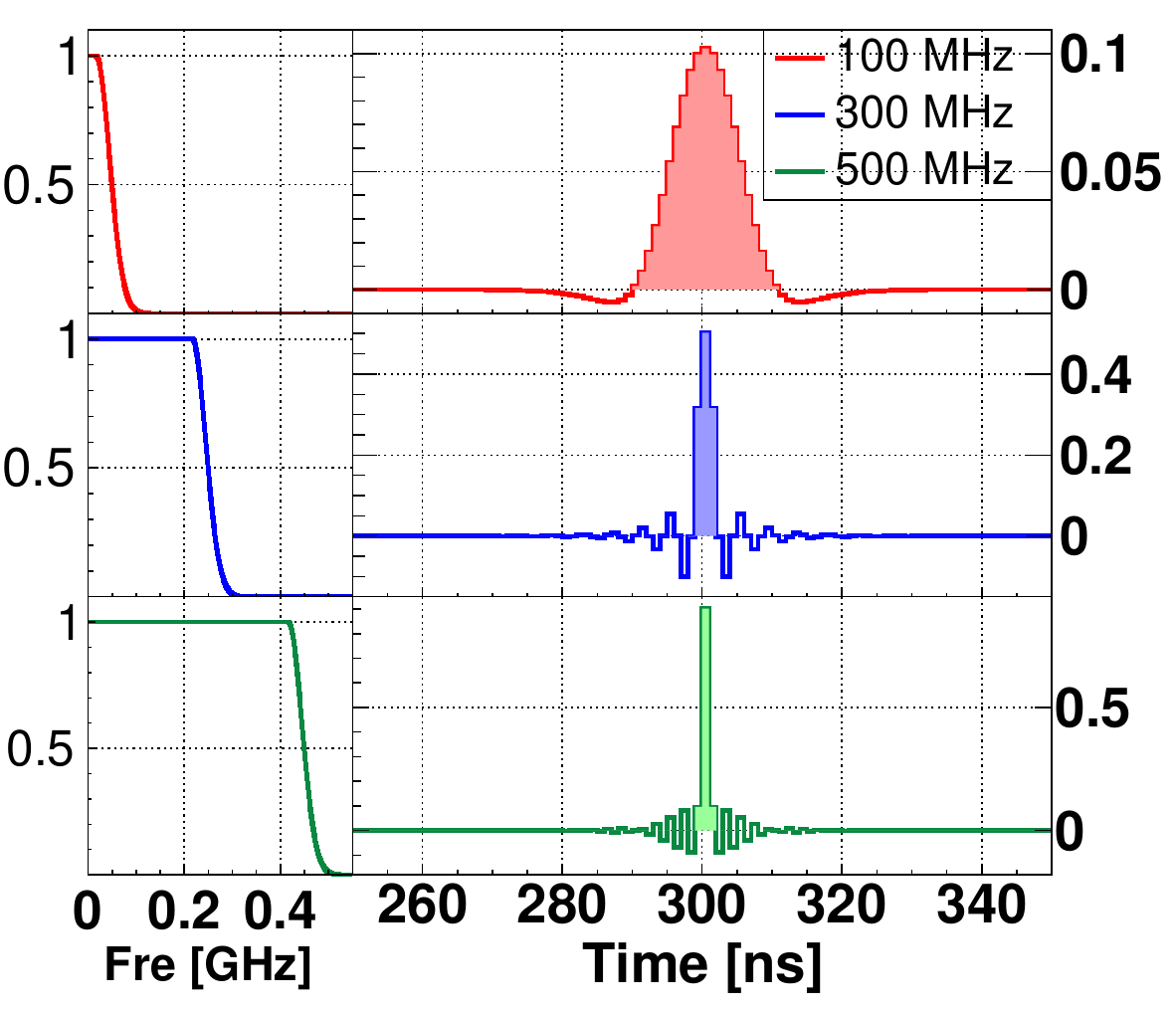}
\caption{\label{fig:dayabayc:c}The custom-defined low-pass filter.}
\end{subfigure}
\begin{subfigure}{0.49\textwidth}
\includegraphics[width=0.95\linewidth]{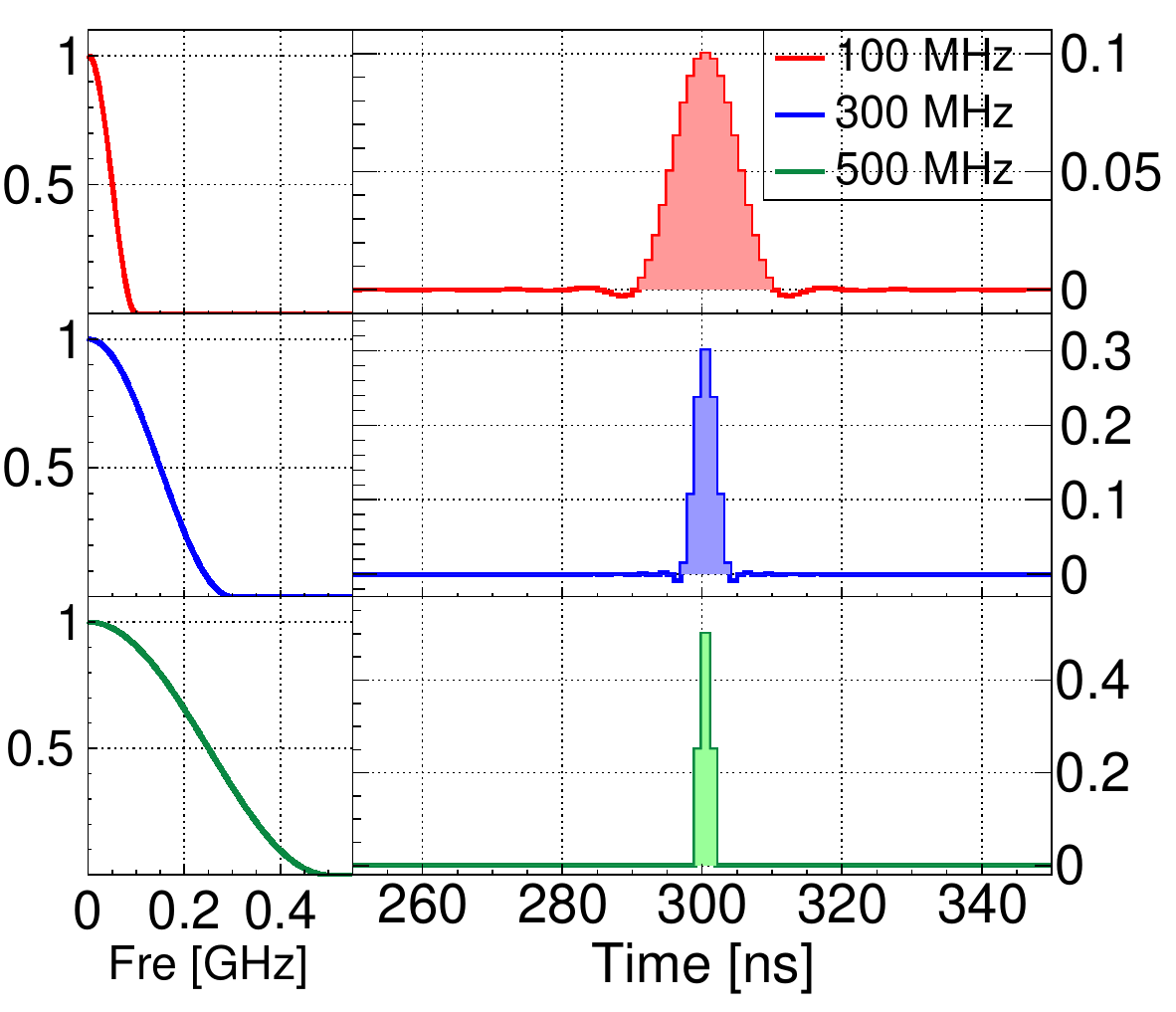}
\caption{\label{fig:cosined:d}The cosine filter.}
\end{subfigure}
\caption{\label{fig:test} The reconstruction results of a 1 PE hit signal with different configurations of noise filters. (a) The three ideal low-pass filters only retain components below their cutoff frequencies (100, 200 and 500~MHz). (b) The filter response of the three Gaussian filters follows a Gaussian function, with a central value of 0 and sigma values equal to one-third of 100, 300, and 500~MHz respectively. There is no further constraint to set the high-frequency response to zero. The cutoff frequencies are defined as three times their respective sigma values. (c) For the three custom-defined low-pass filters, their filter response is equal to 1 at less than 15, 215, and 415~MHz, respectively, and extends to the high-frequency range in the form of a Gaussian function. The three Gaussian functions have the same sigma value, one-third of 85~MHz. 100, 300, and 500~MHZ correspond to their positions of three times sigma, respectively. (d) The filter response of the three cosine filters is a cosine function with angular frequencies of $\frac{\pi}{200}$, $\frac{\pi}{600}$ and $\frac{\pi}{1000}$~MHz$^{-1}$, respectively. And it equals to 0 when greater than 100, 300 and 500~MHz, respectively.}
\end{figure}

An illustrative example is shown in figure~\ref{fig:test} for analysis. We consider a 1 PE hit at 300~ns as the $u(t)$ signal and examine the impact of various filters on the pulse width of the reconstructed hit. The study employed four distinct types of filters, each of which had three different sets of configuration parameters. 

To facilitate a separate examination of filter effects, we temporarily ignore the electronic noise and convolution kernel mismatches. In this case, equation~\ref{eq:deconv_details} can be simplified to equation~\ref{eq:deconv_filterEffect}.

\begin{equation}
\label{eq:deconv_filterEffect}
\begin{gathered}
u_{\rm rec}\left(t \right) = \mathcal{F}^{-1}\left[\mathcal{F}\left[u\left(t \right) \right] \times filter\left(f \right) \right] 
= u\left(t \right) \ast \mathcal{F}^{-1}\left[filter\left(f \right) \right]
\end{gathered}
\end{equation}

In figure~\ref{fig:test}, when comparing different cutoff frequency settings of the same filter, it can be observed that the width of the time domain pulse is inversely correlated with the width of the filter. This correlation can be explained by the time-frequency uncertainty principle. The shapes of reconstructed hits obtained by applying different filters also vary, primarily due to equation~\ref{eq:deconv_filterEffect}. For instance, an ideal low-pass filter acts as a gate signal in the frequency domain and its inverse Fourier transform results in a sampling function. This explains the local ringing around the main peak signal depicted in figure~\ref{fig:ideala:a}. On the other hand, when using a pure Gaussian filter, its inverse Fourier transform remains a Gaussian function resulting in reconstructed hits without ringing as shown in figure~\ref{fig:gausb:b}. Additionally, examples of the custom-defined low-pass filter and the cosine filter are demonstrated in figure~\ref{fig:dayabayc:c} and figure~\ref{fig:cosined:d}, respectively. Their corresponding reconstruction results can also be explained by equation~\ref{eq:deconv_filterEffect}.

\begin{figure}[H]
\centering
\begin{subfigure}{0.49\textwidth}
\includegraphics[width=0.9\linewidth]{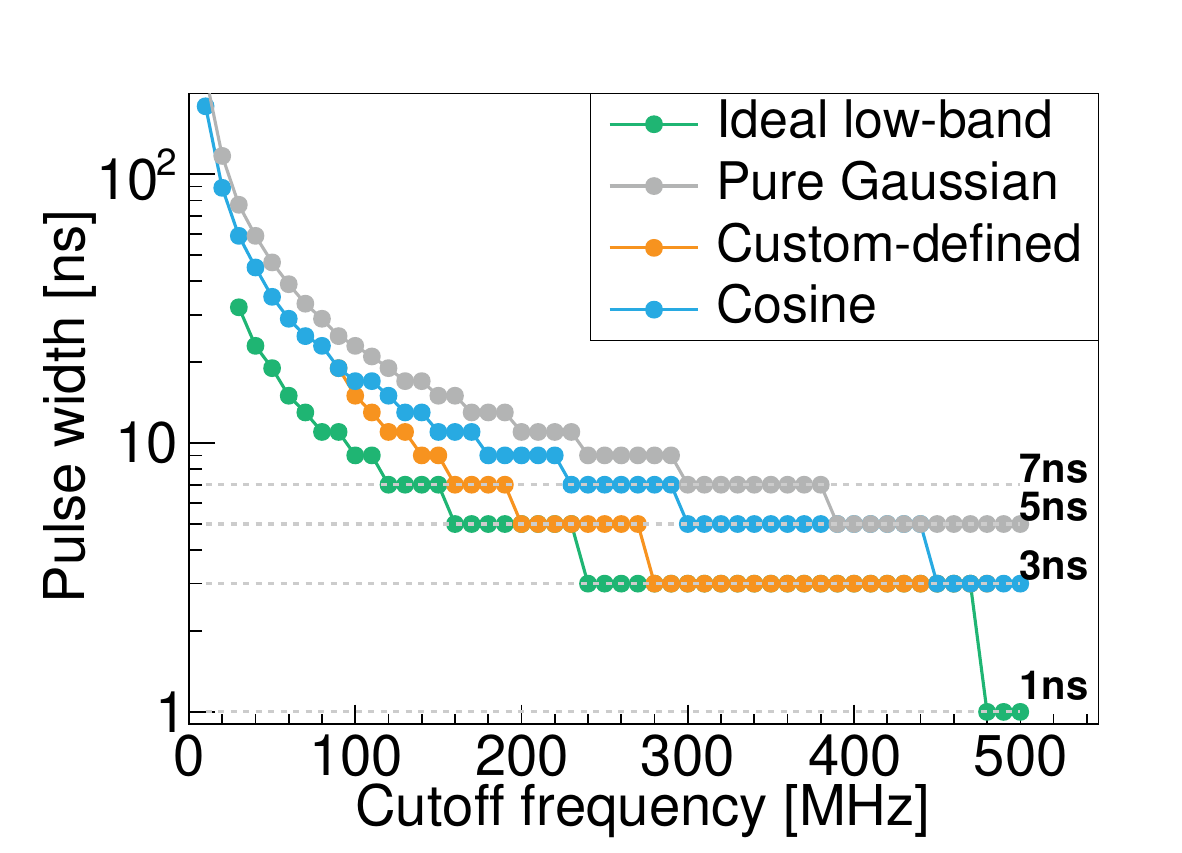}
\caption{\label{fig:width of cut:a}Cutoff frequency and pulse width.}
\end{subfigure}
\begin{subfigure}{0.49\textwidth}
\includegraphics[width=0.9\linewidth]{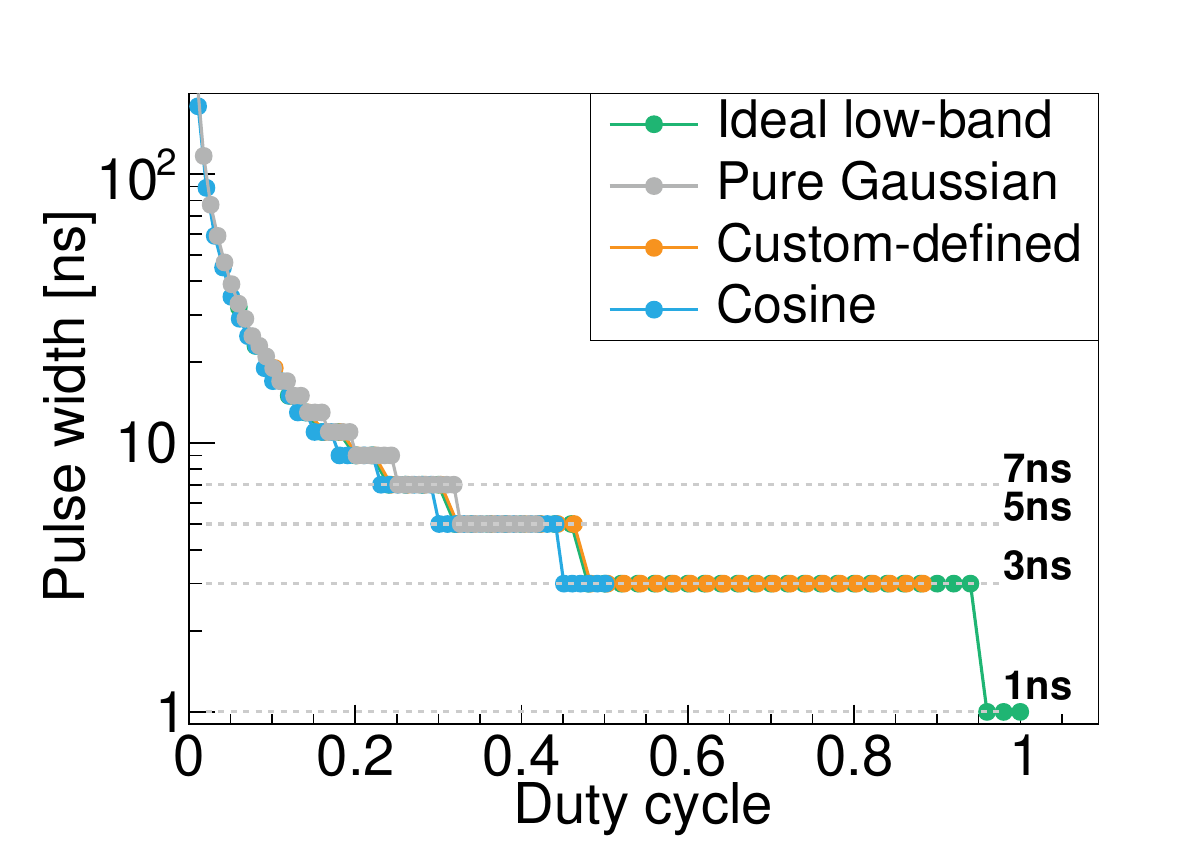}
\caption{\label{fig:duty_cycle:b}Duty cycle and pulse width.}
\end{subfigure}
\caption{\label{fig:cut and empty}The relationship between the pulse width of reconstructed hit and both the cutoff frequency and duty cycle. The results in the figure do not take into account the effects of convolution kernel mismatch and electronic noise.}
\end{figure}

Additionally, the relationship between the pulse width of reconstructed hits and both the cutoff frequency and duty cycle (which refers to the proportion of enclosed area in the filter) is illustrated in figure~\ref{fig:cut and empty}. We use four types of filters for detailed analysis and comparison. The impact of the low-pass filter on pulse width demonstrates a decrease in pulse width as the cutoff frequency increases. Similarly, an increase in duty cycle results in a reduction in signal width. This behavior is consistent with the time-frequency uncertainty principle. On the other hand, although different filters exhibit varying influences at identical cutoff frequencies due to shape disparities, they demonstrate consistent agreement when quantified by duty cycle. According to figure~\ref{fig:ideala:a} and figure~\ref{fig:width of cut:a}, in the absence of convolution kernel mismatch and electronic noise, the deconvolution method is capable of reconstructing the waveform into pulses with a width of 1~ns without employing any filter (or utilizing an ideal low-pass filter with a cutoff frequency of 500~MHz). But in reality, both convolution kernel mismatch and high-frequency noise interference have significant impacts on the reconstructed pulse width.

\subsubsection{Summary of the factors affecting the separation of pile-up hits}
\label{sec:summary-pulse-width}

In this section, two main factors affecting the separation of pile-up hits are studied. On one hand, the presence of convolution kernel mismatch slightly increases the pulse width from 1~ns to 2~ns. On the other hand, effective filtering becomes necessary due to high-frequency noise interference. As a result, a pulse width of 1~ns becomes impossible in reconstruction. Next, section~\ref{sec:STFT} demonstrates the optimization of the deconvolution method, which includes selecting noise filters to achieve a reconstructed pulse width of $3\sim5$~ns and accurately estimate its charge.

\subsection{Optimization and charge calculation}
\label{sec:STFT}

To achieve a narrow ($3\sim5$~ns) reconstructed hit, as shown in figure~\ref{fig:cut and empty}, the duty cycle of the filter should not be less than 0.32. Additionally, to avoid interference with charge estimation, the time domain response of the filter should not have a significant ringing component. Therefore, two filters have been investigated to the subsequent reconstruction: a cosine filter with a cutoff frequency of 450~MHz and a pure Gaussian filter with a cutoff frequency of 400~MHz. The former is expected to result in a reconstructed pulse width of 3~ns, while the latter is expected to correspond to 5~ns. However, high-frequency noise cannot be effectively removed in these filter configurations, requiring additional consideration for accurate charge calculation. In this section, the frequency spectrum characteristics of the signal and noise in the reconstruction process are investigated, followed by optimization of the selection of signal pulses and charge calculation. 

\subsubsection{The evolutionary behavior of the frequency spectrum}
\label{sec:evolutionary}

\begin{figure}[H]
\centering
\begin{subfigure}{0.49\textwidth}
\includegraphics[width=0.9\linewidth]{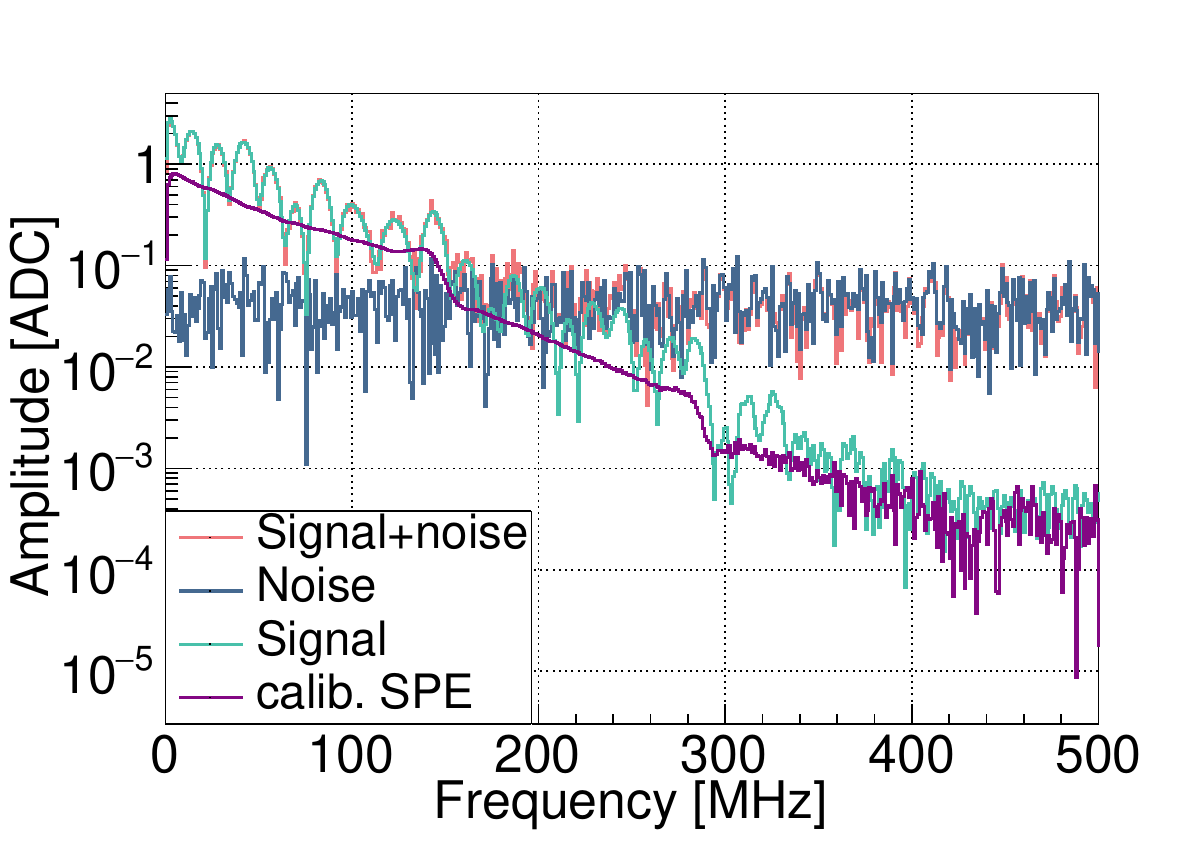}
\caption{\label{fig:component_compare:a}The frequency spectrum of PMT waveform.}
\end{subfigure}
\begin{subfigure}{0.49\textwidth}
\includegraphics[width=0.9\linewidth]{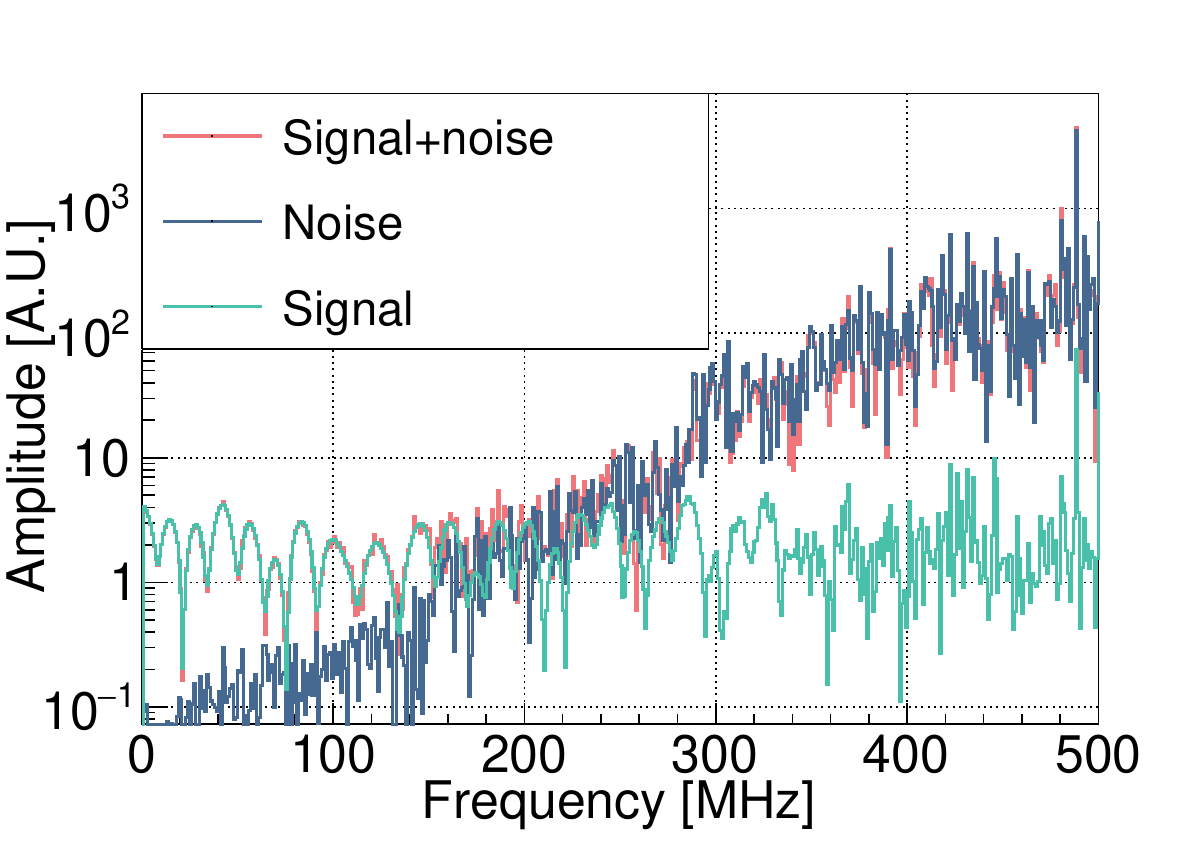}
\caption{\label{fig:component_deconv:b}The deconvolution spectrum of PMT waveform.}
\end{subfigure}
\begin{subfigure}{0.49\textwidth}
\includegraphics[width=0.9\linewidth]{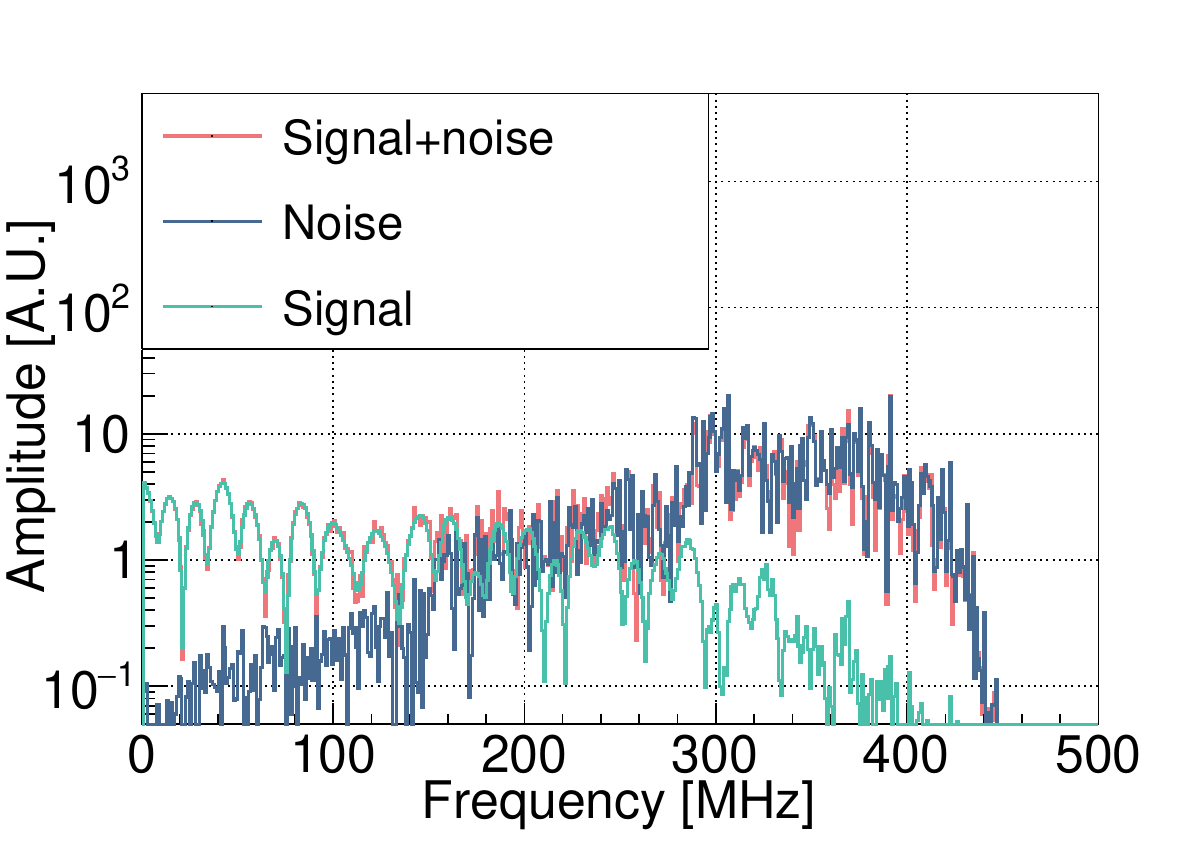}
\caption{\label{fig:component_deconv_fre:c}The deconvolution spectrum after filtering with the cosine filter.}
\end{subfigure}
\begin{subfigure}{0.49\textwidth}
\includegraphics[width=0.9\linewidth]{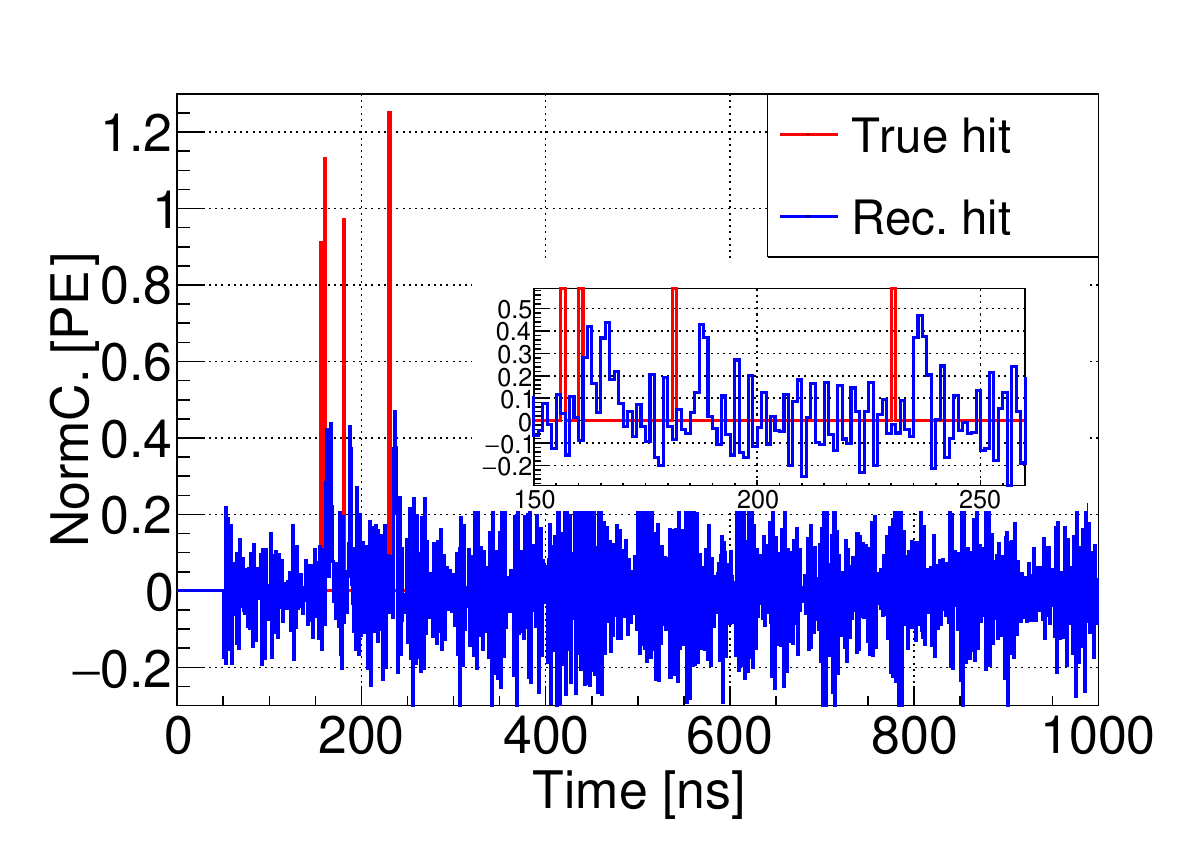}
\caption{\label{fig:component_deconv:d}The reconstruction result ($u_{\rm rec}(t)$) after filtering with the cosine filter.}
\end{subfigure}
\begin{subfigure}{0.49\textwidth}
\includegraphics[width=0.9\linewidth]{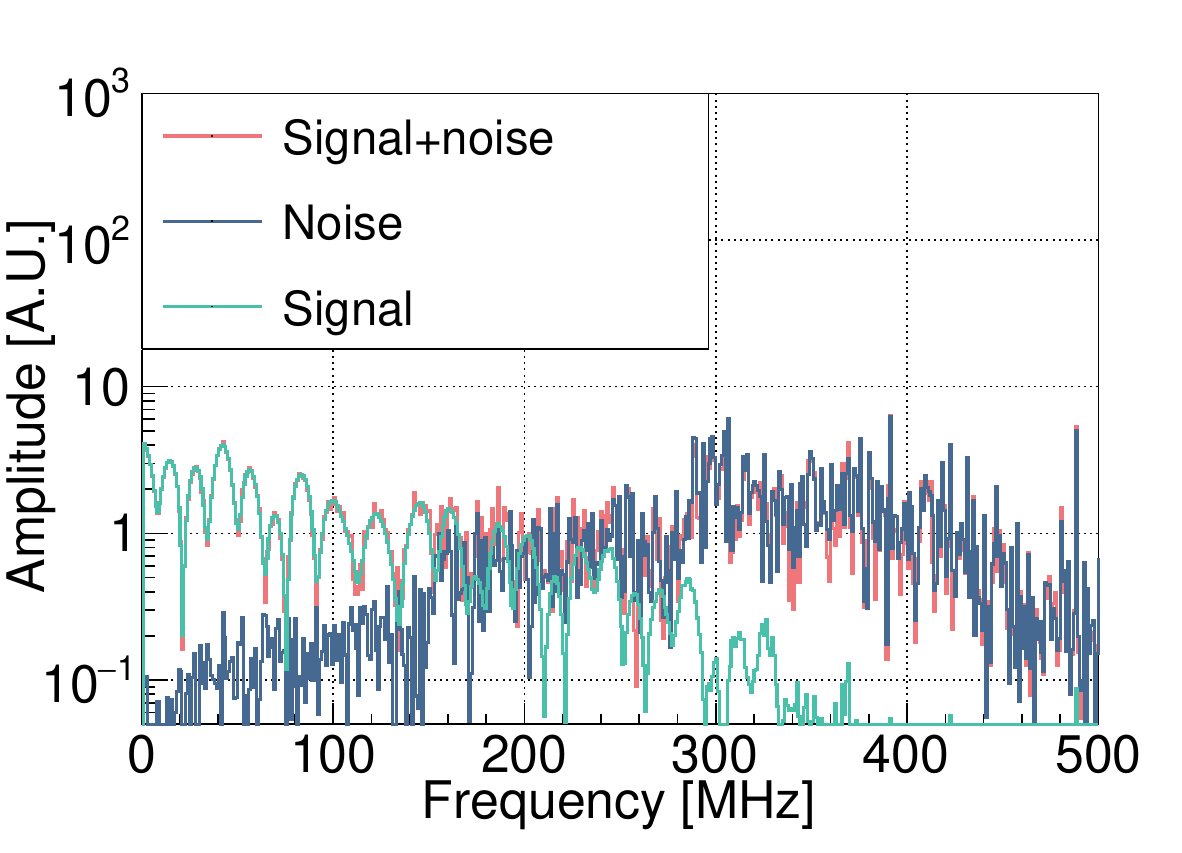}
\caption{\label{fig:component_deconv_gaus:e}The deconvolution spectrum after filtering with the pure Gaussian filter.}
\end{subfigure}
\begin{subfigure}{0.49\textwidth}
\includegraphics[width=0.9\linewidth]{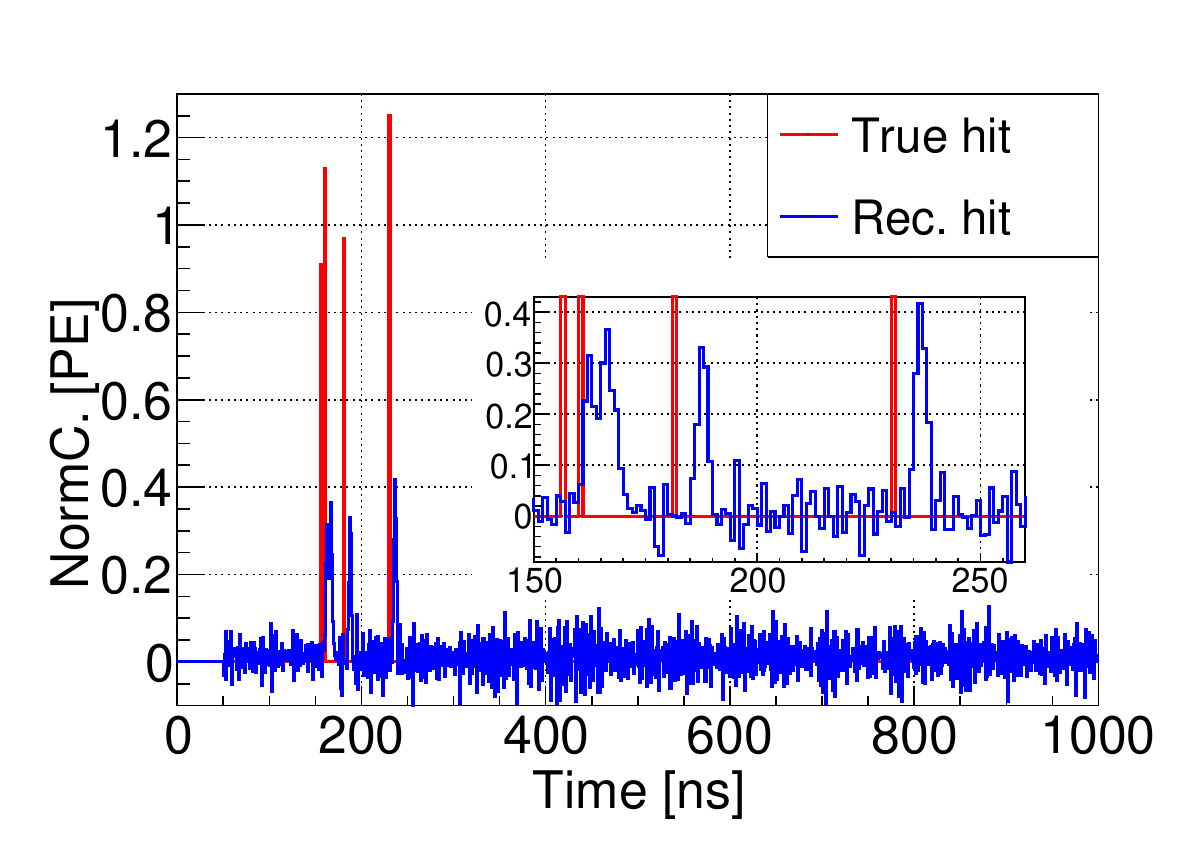}
\caption{\label{fig:component_deconv:f}The reconstruction result ($u_{\rm rec}(t)$) after filtering with the pure Gaussian filter.}
\end{subfigure}
\caption{\label{fig:component_deconv_evolution}The evolutionary behavior of the frequency spectrum during the reconstruction process. The waveform in this example is from figure~\ref{fig:Waveform_Charge}. The frequency spectra of the complete waveform (“Signal + Noise”), the signal (“Signal”), the electronic noise (“Noise”), and the calibrated SPE waveform (“Calib.~SPE”) are shown in (a). Their deconvoluted spectra are shown in (b). The consine filter with a cutoff frequency of 450~MHz is applied in (c) and (d), and the Gaussian filter with a cutoff frequency of 400~MHz is applied in (e) and (f). During the deconvolution, the starting time of the SPE template waveform is fixed at 50~ns, resulting in an earlier occurrence (50~ns) of the deconvoluted pulse compared to the raw waveform. After baseline correction, the reconstructed result is shifted by 50~ns to the right for easier comparison, and the baseline for the first 50~ns is set to zero. This explains why both (d) and (f) have a baseline of zero for the first 50~ns.}
\end{figure}

The waveform in figure~\ref{fig:Waveform_Charge} is utilized as an illustrative example in figure~\ref{fig:component_deconv_evolution} to analyze and compare the evolutionary behavior of the frequency spectrum during the reconstruction process. The frequency spectrum of the complete waveform (“Signal + Noise”), the signal (“Signal”), the electronic noise (“Noise”), and the calibrated SPE waveform (“Calib.~SPE”) is displayed in figure~\ref{fig:component_compare:a}. The signal frequency was concentrated to less than 300~MHz, and white noise was applied in the PMT waveform simulation for electronic noise. Then, each spectrum in figure~\ref{fig:component_compare:a} is divided by the frequency response of the calibrated SPE waveform to obtain the deconvoluted spectra shown in figure~\ref{fig:component_deconv:b}. 

It can be found that the power of the high-frequency noise is relatively amplified after this operation. To achieve a narrow ($3\sim5$~ns) reconstructed hit and reduce the local ringing around the main peak, the cosine filter and pure Gaussian filter are subsequently applied to each deconvoluted spectrum depicted in figure~\ref{fig:component_deconv:b}, resulting in figure~\ref{fig:component_deconv_fre:c} and figure~\ref{fig:component_deconv_gaus:e}. Evidently, since we need a filter with a large bandwidth to achieve a narrow reconstructed hit, the filter configuration used cannot effectively suppress high-frequency noise. As a result, the filtered high-frequency noise still contributes significantly to the frequency spectrum. Finally, as shown in figure~\ref{fig:component_deconv:d} and figure~\ref{fig:component_deconv:f}, in the time-domain response obtained after Inverse DFT, the reconstructed hit in figure~\ref{fig:component_deconv:d} is accompanied by a lot of high-frequency noise which poses challenges for correctly locating the pulse corresponding to the signal and performing accurate charge estimation. Figure~\ref{fig:component_deconv:f} employs a low-pass filter with a smaller duty cycle, leading to clearer reconstructed hits with fewer high-frequency noise. However, the width of each reconstructed hit is approximately 5~ns, which is relatively broader compared to figure~\ref{fig:component_deconv:d}. On the other hand, the amplitude of each reconstructed hit in both figure~\ref{fig:component_deconv:d} and figure~\ref{fig:component_deconv:f} is lower than that of the corresponding true hit due to the differing pulse widths but similar pulse areas (charge [PE]).

\subsubsection{Optimization based on STFT}
\label{sec:Optimization-STFT}

The pulse corresponding to the signal is predominantly concentrated in the low-frequency region and within a short time window, while the high-frequency noise is uniformly distributed throughout the entire time window. So, it is possible to achieve a high SNR locally and perform signal pulse selection. Considering that FFT can only display the overall frequency information, it lacks the ability to provide time-localized frequency information. Therefore, we resort to using the Short-Time Fourier Transform (STFT) to locate the signal pulse. STFT is a mathematical tool that converts signals between the time and frequency domains. The basic idea is to segment the signal, treating each segment as approximately stationary, and then apply the Fourier Transform to each segment to obtain time-localized frequency information~\cite{Grchenig2001_STFT}.

The main steps of STFT analysis and signal pulse selection are as follows:

(1) The reconstruction result of FFT ($u_{\rm rec}(t)$, as shown in figure~\ref{fig:component_deconv:d}) is sliced along the time axis, with time overlap between each segment keeping the same number of points per waveform regardless of the window size chosen. The window size and step length can be adjusted as needed; in this study, a 20~ns window and a 1~ns step were applied.

(2) A window function is applied to each segment to reduce spectral leakage. Commonly used window functions include the Gaussian window, Hamming window, Hann window, and rectangular window. In this paper, the Gaussian window function is used.

\begin{figure}[H]
    \centering
    \begin{tikzpicture}
        \node at (-1.5, 2) {\includegraphics[width=0.4\textwidth]{rec_figure/cosine_460_back.pdf}};
        \node at (-1.3, 3.3) {\small (a)};

        \node at (-1.5, -2.5) {\includegraphics[width=0.4\textwidth]{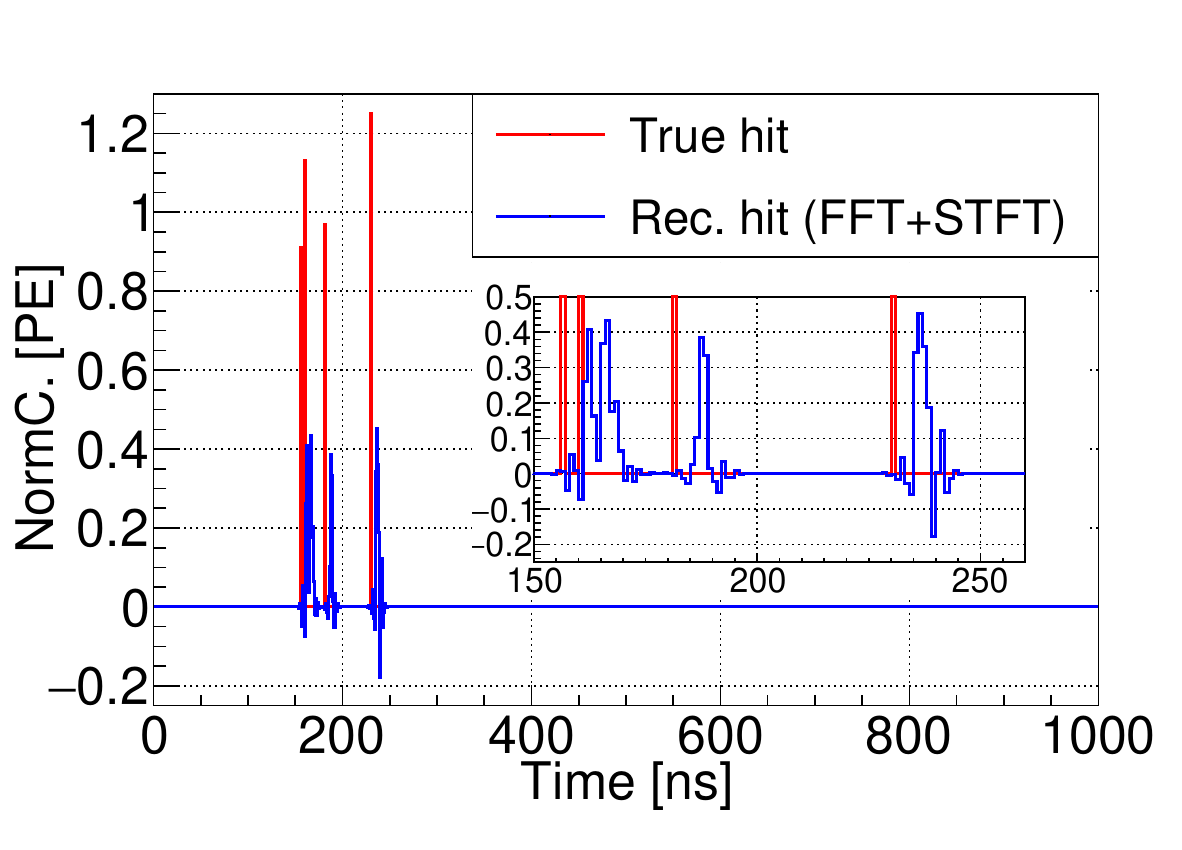}};
        \node at (-1.3, -3.65) {\small (d)};

        \node at (5.5, 2) {\includegraphics[width=0.4\textwidth]{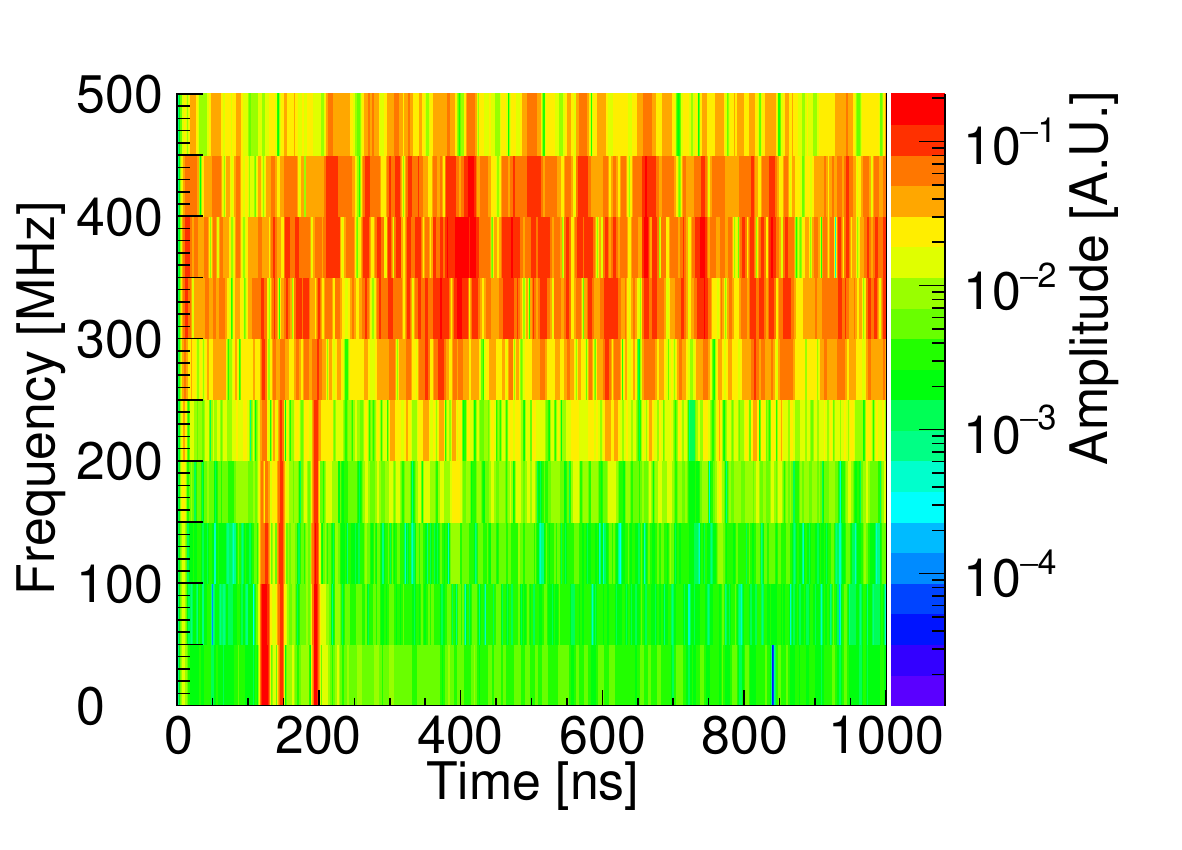}};
        \node at (5.5, 1) {\small (b)};

        \node at (5.5, -2.5) {\includegraphics[width=0.4\textwidth]{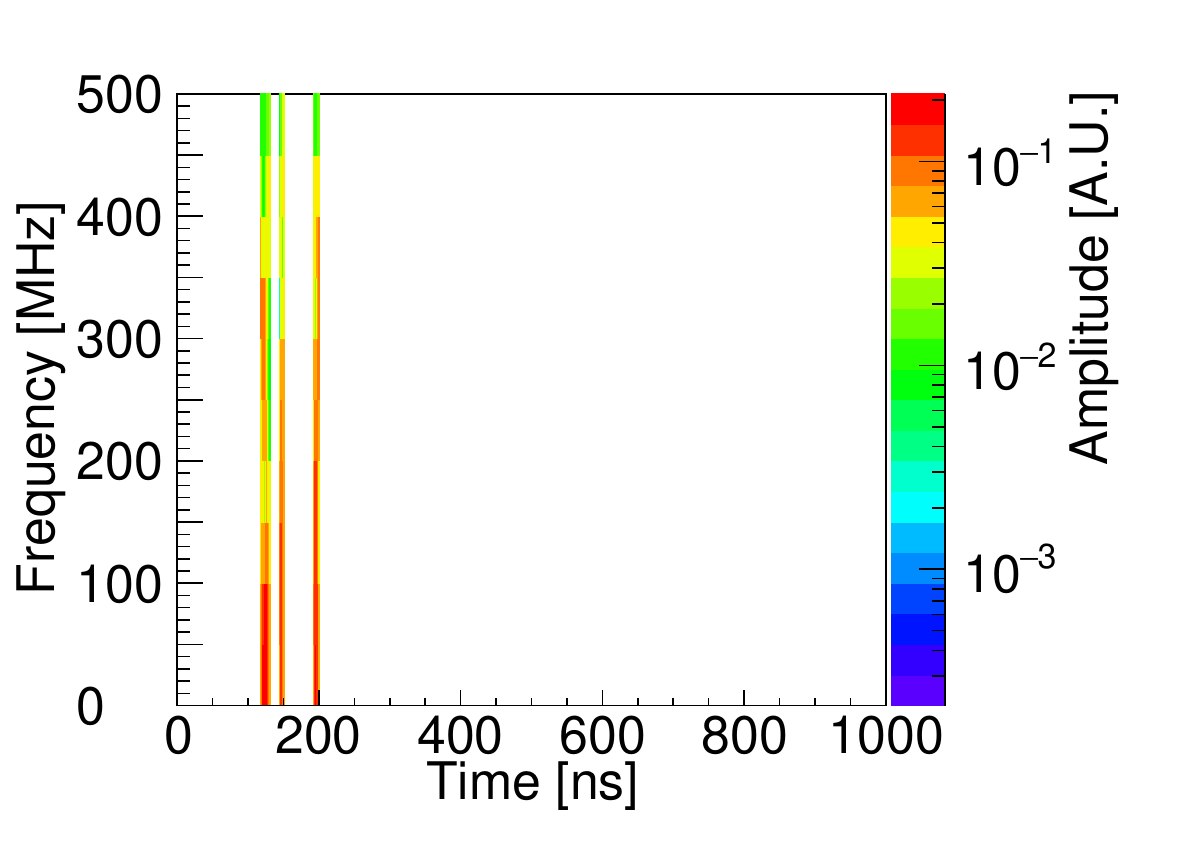}};
        \node at (5.5, -3.65) {\small (c)};

        \draw[->,black, very thick](1.4, 2.0) -- (2.2, 2);
        \draw[->,black, very thick](5.5, -0.1) -- (5.5, -0.5);
        \draw[->,black, very thick](2.2, -2.5) -- (1.4, -2.5);
        \node at (1.8, 2.25) {STFT};
        \node at (6.5, -0.3) {Threshold};
        \node at (1.8, -2.25) {iSTFT};

        \draw[->,black, very thick](4.5, -1.885) -- (7.2, -1.885);
        \node at (5.85, -1.2) {Threshold:};
        \node at (5.85, -1.65) {$ \geq 0.02 \; [\text{A.U.}]$};

    \end{tikzpicture}
    \caption{\label{fig:STFT}Flow chart of STFT analysis and signal pulse selection. In arrow order, they are: (a) The reconstruction result ($u_{\rm rec}(t)$) of FFT after filtering with the cosine filter, which is also shown in figure~\ref{fig:component_deconv:d}. (b) The time-frequency spectrum of (a) obtained by STFT. (c) The time-frequency spectrum of the signal pulse candidate. The color corresponds to the amplitude of the frequency power. (d) The reconstructed hits ($u^{\rm FFT+STFT}_{\rm rec}(t)$) obtained by inverse STFT.}
\end{figure}

(3) A Fourier transform is applied to each segment, and the results are organized over time to obtain the time-frequency representation of the signal. As shown in figure~\ref{fig:STFT}(b), the low-frequency signal mainly appears in a few time periods of about 150 to 200~ns. This time-localized frequency feature enables precise searching for signal pulses.

(4) In the frequency range below 50~MHz, a threshold is applied to the amplitude of the frequency power to extract the time-frequency spectrum of the signal pulse candidates (figure~\ref{fig:STFT}(c)); the threshold was set to 0.02 A.U.

(5) The reconstructed hits ($u^{\rm FFT+STFT}_{\rm rec}(t)$) can be obtained after converted to the time domain with inverse STFT (iSTFT), as shown in figure~\ref{fig:STFT}(d).

After the aforementioned operations, it can be observed that the STFT signal-finding method is applicable in cases of poor SNR. These operations address the shortcomings in noise reduction encountered when using the cosine filter to achieve a narrow reconstructed hit and reduce the local ringing around the main peak (as illustrated in figure~\ref{fig:component_deconv_fre:c} and figure~\ref{fig:component_deconv:d}). Additionally, the reconstructed hit obtained through STFT analysis becomes more prominent, thereby facilitating precise charge reconstruction. Based on the experience of FADC waveform reconstruction in Daya Bay, this paper also integrates pulses within the selected integral region to obtain the reconstructed charge. In order to select the signal hit candidate in figure~\ref{fig:STFT}(d), a threshold of 0.2 is applied to each pulse amplitude, followed by integrating the peak region and its preceding and subsequent amplitudes for 5~ns for each signal hit candidate to provide the reconstructed charge. The performance of our optimized deconvolution reconstruction algorithm is shown and discussed in the next section.

\subsection{Performance and discussion}
\label{sec:Performance}

\begin{figure}[H]
\centering
\begin{subfigure}{0.99\textwidth}
\includegraphics[width=0.99\linewidth]{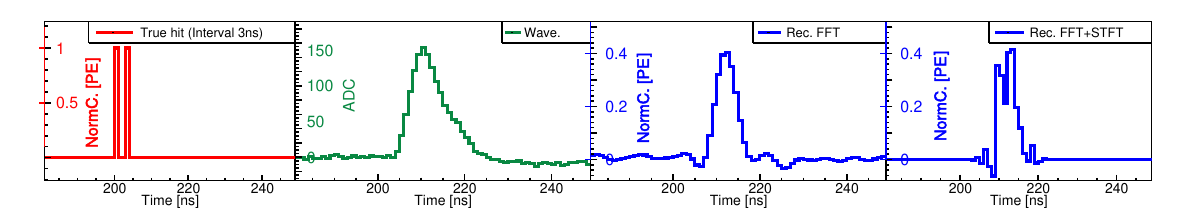}
\caption{\label{fig:pile_up_3ns}Two hits separated by 3~ns. }
\end{subfigure}
\begin{subfigure}{0.99\textwidth}
\includegraphics[width=0.99\linewidth]{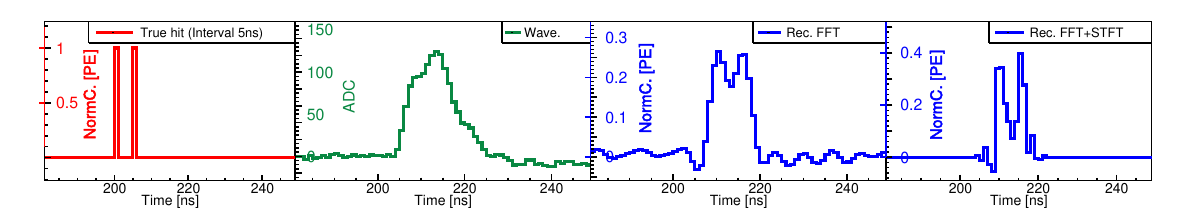}
\caption{\label{fig:pile_up_5ns}Two hits separated by 5~ns. }
\end{subfigure}
\begin{subfigure}{0.99\textwidth}
\includegraphics[width=0.99\linewidth]{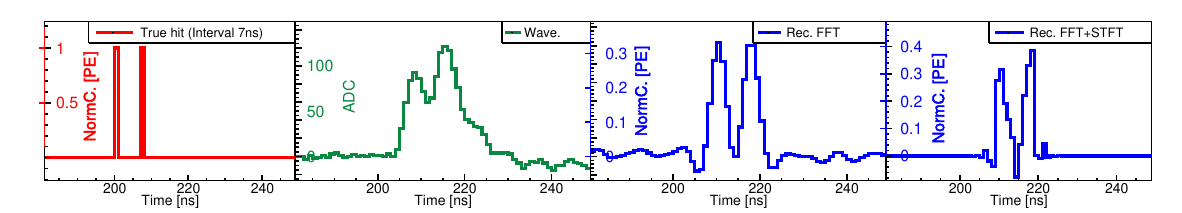}
\caption{\label{fig:pile_up_7ns}Two hits separated by 7~ns. }
\end{subfigure}
\begin{subfigure}{0.99\textwidth}
\includegraphics[width=0.99\linewidth]{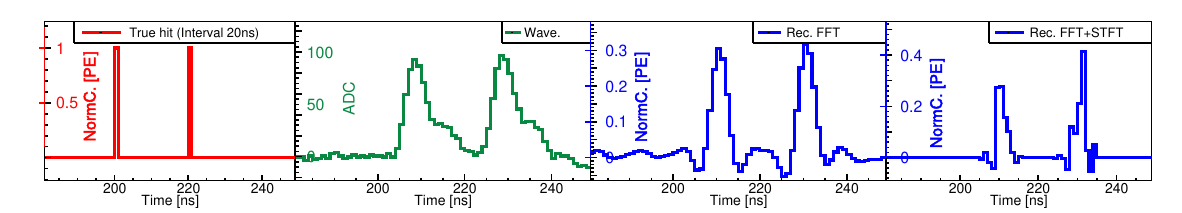}
\caption{\label{fig:pile_up_20ns} Two hits separated by 20~ns.}
\end{subfigure}
\caption{\label{fig:component_deconv}The reconstruction results of two hits in the case of separating by 3~ns, 5~ns, 7~ns and 20~ns. The pile-up hits can be more clearly identified after reduced the width of the reconstructed hits.}
\end{figure}

After algorithm optimization, we investigated its performance with simulated waveforms. In order to intuitively compare the resolution of pile-up hits, we generated two isolated hits and set their time intervals to 3, 5, 7, and 20~ns, respectively. We then reconstructed their waveforms using the deconvolution algorithm before (FFT) and after optimization (FFT+STFT), as shown in figure~\ref{fig:component_deconv}. As introduced in section~\ref{sec:Investigation-optimization}, this paper mainly uses the pulse width of a single hit to characterize the algorithm's ability to separate pile-up hits. Thus, as shown in figure~\ref{fig:component_deconv}, the pile-up hits can be more clearly identified after reduced the width of the reconstructed hits. For example, in figure~\ref{fig:pile_up_5ns}, when reconstructed using the optimized algorithm, two hits with an interval of 5~ns can be recognized effectively.

\begin{figure}[H]
\centering 
\includegraphics[width=0.49\textwidth]{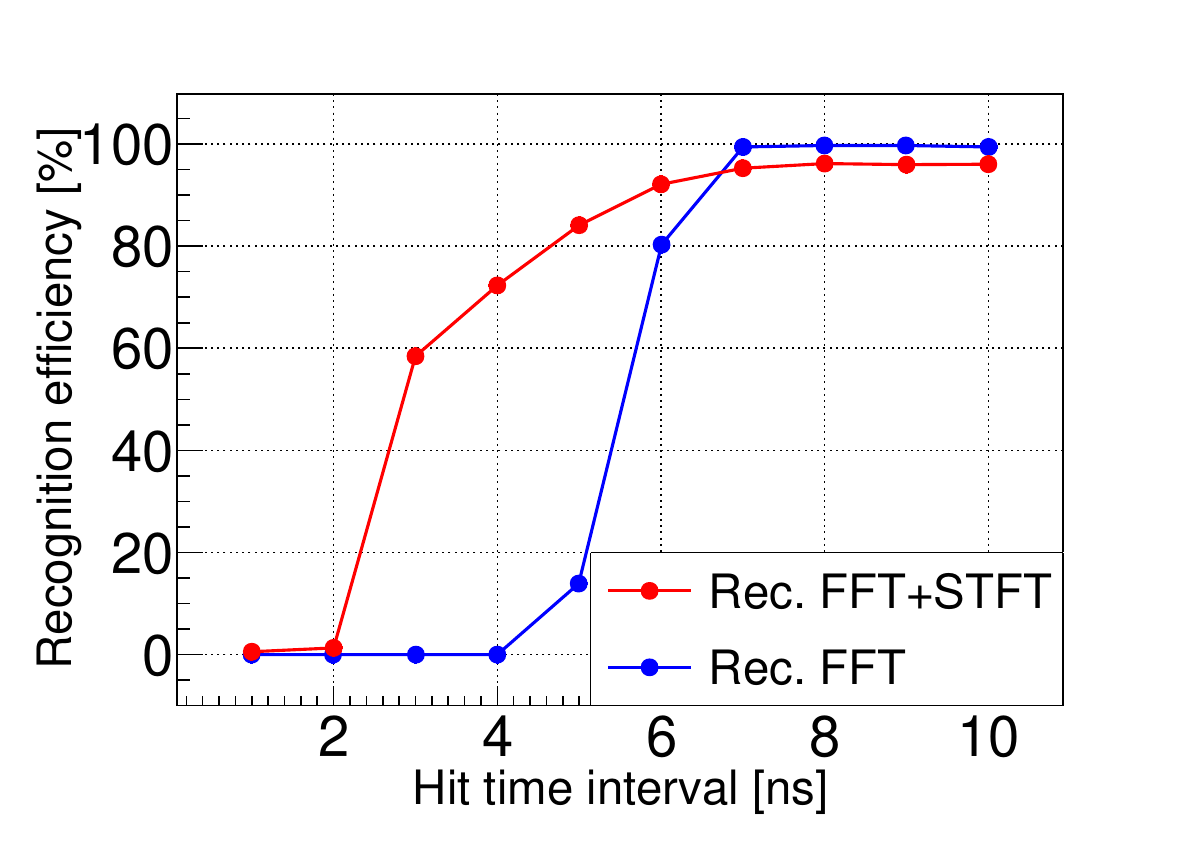}
\caption{\label{fig:rec_efficiency} The recognition efficiency of two reconstruction methods for detecting two hits with varying time intervals. The enhanced method (Rec. FFT+STFT) significantly enhances the capability to identify pile-up hits with time intervals between 3 to 6~ns. However, the wider noise filter introduces more high-frequency noise, which may affect the identification of narrower reconstructed hits, leading to potential misjudgments. Consequently, the recognition efficiency for hits with time intervals of 7 to 10~ns is slightly reduced by about 3\% compared with the previous method (Rec. FFT).}
\end{figure}

\begin{figure}[H]
\centering 
\includegraphics[width=0.99\textwidth]{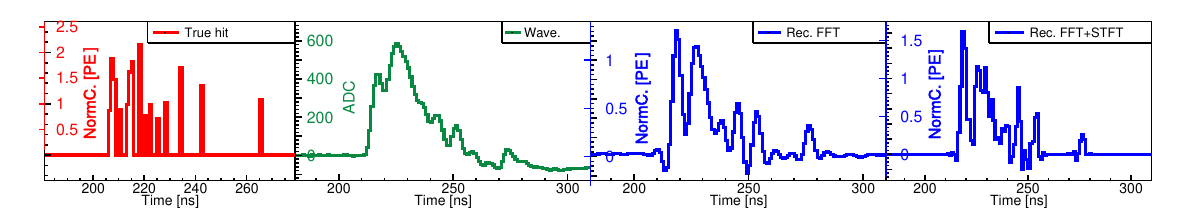}
\caption{\label{fig:pile_up_MC} A waveform with 18 true hits and its reconstructed hits. The total charge of these hits is 19.31~PEs after taking into account the charge dispersion of individual photoelectrons, and the reconstructed charge is 18.81~PEs and 18.61~PEs for the enhanced method (Rec. FFT+STFT) and the previous method (Rec. FFT), respectively.}
\end{figure}

To evaluate the ability of the reconstruction methods to recognize pile-up hits, more simulated waveforms consisting of two hits with time intervals of 1 to 10 ns were generated, with 10000 simulated waveforms for each case. Each hit included the complete simulation process described in section~\ref{sec:sample} and was then reconstructed. After reconstruction, we established a simple recognition criterion to calculate and verify the efficiency of tagging the two hits using the two reconstruction methods. In the recognition criterion, candidate pulses that exceed the amplitude threshold will undergo further selection, combination, or elimination based on their width and area (charge). Only isolated pulses with both left and right edges returning to or nearly reaching the baseline will be counted as valid hits. If after these operations, exactly two hits can be successfully identified, and the deviation of their hit times from the true simulated times does not exceed 2~ns, it is considered a successful identification. The recognition criterion employed in this paper only serve as a tool for relative performance comparisons, and it is possible that higher recognition efficiency could be attained through alternative approaches such as machine learning. After analyzing simulated waveforms with varying time intervals, the results presented in figure~\ref{fig:rec_efficiency} are obtained. After optimization, the capability to identify pile-up hits with time intervals ranging from 3~ns to 6~ns is significantly enhanced, primarily attributed to the reduced pulse width and diminished local ringing around the main peak. For the case with a time interval of 7 to 10~ns, the recognition efficiency is approximately 3\% lower than that of the previous reconstruction method, mainly due to the use of the wider filter that leaves more high-frequency noise. Additionally, the narrower reconstructed hit is more susceptible to high-frequency noise during identification. Figure~\ref{fig:pile_up_MC} shows a waveform with 18 true hits and its reconstructed hits, the pile-up effect is more serious, but it can still be accurately reconstructed, and the optimized method is better for the separation of the pile-up hits.

\begin{figure}[H]
\centering
\includegraphics[width=0.99\linewidth]{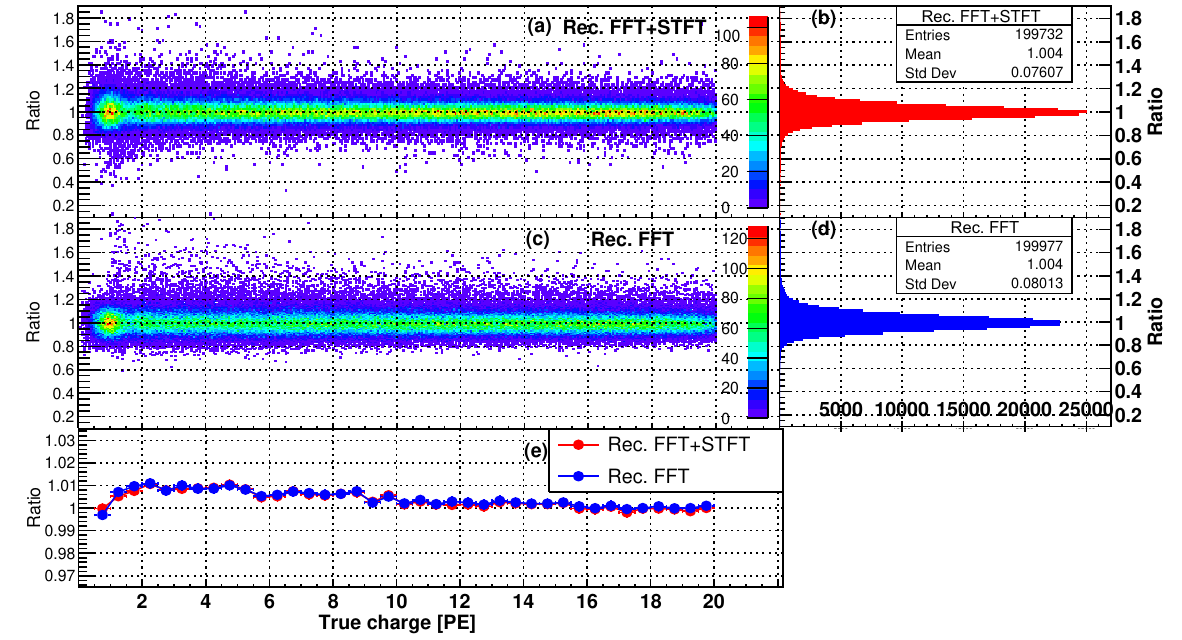}
\caption{\label{fig:STFT_unline_20}The comparison of the reconstructed charge and the true charge. The results of the enhanced method (Rec. FFT+STFT) and the previous method (Rec. FFT) are shown in (a) and (c), respectively. The X-axis is MC true charge, and the Y-axis is the ratio between the reconstructed charge and the MC charge. The color in (a) and (c) corresponds to the number of events in a certain bin. (b) and (d) show the overall distributions of the ratios. In (e), the red and the blue dots correspond to “Rec. FFT+STFT” and “FFT” respectively, and these dots indicate the averaged ratio at different PEs. The residual nonlinearity of both method is about 1\%.}
\end{figure}

The charge reconstruction results of 200,000 simulated waveforms (as introduced in section~\ref{sec:sample}) are presented in figure~\ref{fig:STFT_unline_20}, demonstrating the effectiveness of our optimization approach. As a comparison, the reconstruction results obtained using the previous reconstruction method with the same simulated sample are also presented. The enhanced method improves the discrimination capability for pile-up hits while keeping an accurate reconstructed charge in the range of $0\sim20$~PEs, with residual nonlinearity of about 1\%. We investigated the few cases (the blue dots) in which the reconstructed charge deviated significantly from the true charge (ratios greater than 1.5 or less than 0.5) and found that this was mainly due to the threshold applied to the amplitude of the frequency power when we try to extract the time-frequency spectrum of the signal after STFT. As shown in figure~\ref{fig:STFT}, we adopted a 0.02 threshold from (b) to (c). This threshold ensures that small charge signals (less than 0.5 PEs) can be detected, but it may also introduce some false signals when the baseline fluctuates sharply, leading to the reconstruction deviation shown in figure~\ref{fig:STFT_unline_20}(a). As illustrated in figure~\ref{fig:STFT_unline_20}(b), instances of significant deviations are indeed infrequent, thus the overall reconstruction remains reliable. This dispersion can be further reduced by improving the selection of the threshold value according to the actual situation, but it needs to take into account the detection of small charges. For example, we have tried adjusting the threshold to 0.03, in which case the standard deviation shown in figure~\ref{fig:STFT_unline_20}(b) can be reduced from $\sim$0.076 to $\sim$0.065 and there are no more reconstructed results with ratios (figure~\ref{fig:STFT_unline_20}(a)) greater than 1.5, but there will be a few signals with small charges that cannot be detected. 

\section{Summary}
\label{sec:summary}

PMT is widely used in neutrino and dark matter experiments. The reconstruction of PMT waveforms is the basis of accurate measurement and physical detection. In recent years, there have been some new developments and demands for PMT waveform reconstruction, including precise time and charge (or photoelectron count) reconstruction, adaptability to wide dynamic ranges, and enhanced discrimination of pile-up hits. The article focuses on the timing separation capability for pile-up hits as well as accurate charge reconstruction. The deconvolution reconstruction method is successfully applied in the waveform reconstruction of the FADC system in the Daya Bay experiment. However, due to the wide pulse width of the reconstructed hit, it poses a challenge for further separation of pile-up hits, with a separation capability of approximately 7 to 10~ns. After a detailed study, it is found that both the mismatch of convolution kernel and the use of noise filters affect the pulse width of reconstructed hits, and the latter dominates. By exploring the relationship between filter configuration and pulse width, we have successfully optimized the design of the noise filter. As a result, the reconstructed pulse widths for the cosine filter with a cutoff frequency of 450~MHz and the pure Gaussian filter with a cutoff frequency of 400~MHz can achieve 3~ns and 5~ns respectively. In addition, the deconvolution ringing is greatly reduced by choice of appropriate filters. Finally, STFT is used for signal pulse selection and charge reconstruction in the case of low SNR caused by the wide band filter. After our optimization, the enhanced deconvolution method can identify hits with interval greater than 3 or 5~ns and accurately reconstruct the charge of the PMT waveform with a nonlinearity of about 1\% in the range of 0 to 20~PEs. The research results of this paper are expected to be further applied to accurate PMT waveform reconstruction, the development of PE counting technology and particle identification study.

\acknowledgments

This work was supported by National Key R\&D Program of China No. 2023YFA1606103, the Guangxi Science and Technology Program (No. GuiKeAD21220037), and the Program of Bagui Scholars Program (XF).

\bibliographystyle{unsrt}
\bibliography{ref}

\begin{thebibliography}{10}

\bibitem{SNO:1999crp}
J.~Boger et~al.
\newblock {The Sudbury neutrino observatory}.
\newblock {\em Nucl. Instrum. Meth. A}, 449:172--207, 2000.

\bibitem{FUKUDA2003418}
The Super-Kamiokande Collaboration.
\newblock The super-kamiokande detector.
\newblock {\em Nuclear Instruments and Methods in Physics Research Section A: Accelerators, Spectrometers, Detectors and Associated Equipment}, 501(2):418--462, 2003.

\bibitem{KamLAND:2004mhv}
T.~Araki et~al.
\newblock {Measurement of neutrino oscillation with KamLAND: Evidence of spectral distortion}.
\newblock {\em Phys. Rev. Lett.}, 94:081801, 2005.

\bibitem{Borexino:2008gab}
G.~Alimonti et~al.
\newblock {The Borexino detector at the Laboratori Nazionali del Gran Sasso}.
\newblock {\em Nucl. Instrum. Meth. A}, 600:568--593, 2009.

\bibitem{DayaBay:2016ggj}
Feng~Peng An et~al.
\newblock {Measurement of electron antineutrino oscillation based on 1230 days of operation of the Daya Bay experiment}.
\newblock {\em Phys. Rev. D}, 95(7):072006, 2017.

\bibitem{Cheng2017}
Jian-Ping Cheng, Ke-Jun Kang, Jian-Min Li, Jin Li, Yuan-Jing Li, Qian Yue, Zhi Zeng, Yun-Hua Chen, Shi-Yong Wu, Xiang-Dong Ji, and Henry~T. Wong.
\newblock The china jinping underground laboratory and its early science.
\newblock {\em Annual Review of Nuclear and Particle Science}, 67(1):231–251, October 2017.

\bibitem{DarkSide:2018bpj}
P.~Agnes et~al.
\newblock {Low-Mass Dark Matter Search with the DarkSide-50 Experiment}.
\newblock {\em Phys. Rev. Lett.}, 121(8):081307, 2018.

\bibitem{AKERIB2020163047}
The~LZ Collaboration.
\newblock The lux-zeplin (lz) experiment.
\newblock {\em Nuclear Instruments and Methods in Physics Research Section A: Accelerators, Spectrometers, Detectors and Associated Equipment}, 953:163047, 2020.

\bibitem{Li:2021oos}
Ziyuan Li et~al.
\newblock {Event vertex and time reconstruction in large-volume liquid scintillator detectors}.
\newblock {\em Nucl. Sci. Tech.}, 32(5):49, 2021.

\bibitem{Yang:2022din}
Cheng-Feng Yang, Yong-Bo Huang, Ji-Lei Xu, Di-Ru Wu, Hao-Qi Lu, Yong-Peng Zhang, Wu-Ming Luo, Miao He, Guo-Ming Chen, and Si-Yuan Zhang.
\newblock {Reconstruction of a muon bundle in the JUNO central detector}.
\newblock {\em Nucl. Sci. Tech.}, 33(5):59, 2022.

\bibitem{Qian:2021vnh}
Zhen Qian et~al.
\newblock {Vertex and energy reconstruction in JUNO with machine learning methods}.
\newblock {\em Nucl. Instrum. Meth. A}, 1010:165527, 2021.

\bibitem{Zhang:2024okq}
Siyuan Zhang, Yongbo Huang, Miao He, Chengfeng Yang, and Guoming Chen.
\newblock {Sub-GeV events energy reconstruction with 3-inch PMTs in JUNO}.
\newblock 2 2024.

\bibitem{DayaBay:2019yxq}
D.~Adey et~al.
\newblock {Extraction of the $^{235}$U and $^{239}$Pu Antineutrino Spectra at Daya Bay}.
\newblock {\em Phys. Rev. Lett.}, 123(11):111801, 2019.

\bibitem{HUANG201848}
Yongbo Huang, Jinfan Chang, Yaping Cheng, Zhang Chen, Jun Hu, Xiaolu Ji, Fei Li, Jin Li, Qiuju Li, Xin Qian, Soeren Jetter, Wei Wang, Zheng Wang, Yu~Xu, and Zeyuan Yu.
\newblock The flash adc system and pmt waveform reconstruction for the daya bay experiment.
\newblock {\em Nuclear Instruments and Methods in Physics Research Section A: Accelerators, Spectrometers, Detectors and Associated Equipment}, 895:48--55, 2018.

\bibitem{Jiang:2024wph}
Wei Jiang, Guihong Huang, Zhen Liu, Wuming Luo, Liangjian Wen, and Jianyi Luo.
\newblock {Machine-Learning based photon counting for PMT waveforms and its application to the improvement of the energy resolution in large liquid scintillator detectors}.
\newblock 5 2024.

\bibitem{DayaBay:2019fje}
D.~Adey et~al.
\newblock {A high precision calibration of the nonlinear energy response at Daya Bay}.
\newblock {\em Nucl. Instrum. Meth. A}, 940:230--242, 2019.

\bibitem{Zhang_2019}
H.Q. Zhang, Z.M. Wang, Y.P. Zhang, Y.B. Huang, F.J. Luo, P.~Zhang, C.C. Zhang, M.H. Xu, J.C. Liu, Y.K. Heng, C.G. Yang, X.S. Jiang, F.~Li, M.~Ye, and H.S. Chen.
\newblock Comparison on pmt waveform reconstructions with juno prototype.
\newblock {\em Journal of Instrumentation}, 14(08):T08002, aug 2019.

\bibitem{Grassi:2018pxk}
M.~Grassi et~al.
\newblock {Charge reconstruction in large-area photomultipliers}.
\newblock {\em JINST}, 13(02):P02008, 2018.

\bibitem{JUNO:2021vlw}
Angel Abusleme et~al.
\newblock {JUNO physics and detector}.
\newblock {\em Prog. Part. Nucl. Phys.}, 123:103927, 2022.

\bibitem{Chen:2023xhj}
Guo-Ming Chen, Xin Zhang, Ze-Yuan Yu, Si-Yuan Zhang, Yu~Xu, Wen-Jie Wu, Yao-Guang Wang, and Yong-Bo Huang.
\newblock {Discrimination of pp solar neutrinos and $^{14}$C double pile-up events in a large-scale LS detector}.
\newblock {\em Nucl. Sci. Tech.}, 34(9):137, 2023.

\bibitem{Cheng:2023zds}
Jie Cheng, Xiao-Jie Luo, Gao-Song Li, Yu-Feng Li, Ze-Peng Li, Hao-Qi Lu, Liang-Jian Wen, Michael Wurm, and Yi-Yu Zhang.
\newblock {Pulse shape discrimination technique for diffuse supernova neutrino background search with JUNO}.
\newblock {\em Eur. Phys. J. C}, 84(5):482, 2024.

\bibitem{Jetter_2012}
Jetter~Sören et~al.
\newblock Pmt waveform modeling at the daya bay experiment.
\newblock {\em Chinese Physics C}, 36(8):733, aug 2012.

\bibitem{BRUN199781}
Rene Brun and Fons Rademakers.
\newblock Root — an object oriented data analysis framework.
\newblock {\em Nuclear Instruments and Methods in Physics Research Section A: Accelerators, Spectrometers, Detectors and Associated Equipment}, 389(1):81--86, 1997.
\newblock New Computing Techniques in Physics Research V.

\bibitem{GibbsEffect}
J.W. Gibbs.
\newblock {Fourier's Series}.
\newblock {\em Nature}, 59(1539):606, 1899.

\bibitem{Spagnolini2017}
Umberto Spagnolini.
\newblock {\em Statistical Signal Processing in Engineering}.
\newblock Wiley, December 2017.

\bibitem{Sundararajan2024}
Dr.~D. Sundararajan.
\newblock {\em Discrete Fourier Transform}, pages 67--106.
\newblock Springer International Publishing, Cham, 2024.

\bibitem{Heisenberg1927}
W.~Heisenberg.
\newblock Über den anschaulichen inhalt der quantentheoretischen kinematik und mechanik.
\newblock {\em Zeitschrift für Physik}, 43(3–4):172–198, March 1927.

\bibitem{Griffiths_Schroeter_2018}
David~J. Griffiths and Darrell~F. Schroeter.
\newblock {\em Introduction to Quantum Mechanics}.
\newblock Cambridge University Press, 3 edition, 2018.

\bibitem{Boughn2017}
Stephen Boughn.
\newblock Making sense of bell’s theorem and quantum nonlocality.
\newblock {\em Foundations of Physics}, 47(5):640–657, March 2017.

\bibitem{Ream1977}
N.~Ream.
\newblock Discrete-time signal processing.
\newblock {\em Electronics and Power}, 23(2):157, 1977.

\bibitem{Grchenig2001_STFT}
Karlheinz Gr\"{o}chenig.
\newblock {\em Foundations of Time-Frequency Analysis}.
\newblock Birkh\"{a}user Boston, 2001.

\end{thebibliography}

\end{document}